\documentclass[11pt]{article}
\usepackage[a4paper, top=3cm, bottom=3cm, left=2.25cm, right=2.25cm]{geometry}
\usepackage{color}
\usepackage{refcount}
\usepackage{titlesec}
\usepackage{graphicx}
\usepackage{etoolbox}
\usepackage[T1]{fontenc}
\usepackage{float}
\usepackage[toc,page]{appendix}
\usepackage{amssymb}
\usepackage{amsbsy}
\usepackage{dsfont}
\usepackage{lipsum}

\usepackage[numbers,sort&compress]{natbib}
\usepackage{hyperref}

\usepackage{doi}
\usepackage{bookmark}
\hypersetup{
  colorlinks=true,
  linkcolor=blue,
  filecolor=magenta,
  urlcolor=blue,
  citecolor=blue,
}

\usepackage{mathptmx}
\usepackage{braket}
\usepackage{slashed}
\usepackage[dvipsnames]{xcolor}
\usepackage{caption}
\usepackage{amsmath}
\usepackage{subcaption}
\usepackage[utf8]{inputenc}
\usepackage{latexsym}
\usepackage{enumitem}
\usepackage{bbding}
\usepackage{scrextend}
\deffootnote{0.5em}{0em}{\textsuperscript{\thefootnotemark}\,}

\DeclareMathSymbol{\ii}{\mathalpha}{letters}{"10}
\DeclareMathSymbol{\jj}{\mathalpha}{letters}{"11}

\usepackage{scalerel}
\usepackage{euscript}
\usepackage{cancel}
\usepackage{units}
\usepackage{lettrine}
\usepackage{braket}
\usepackage{bm}
\usepackage{booktabs}
\usepackage{cancel}
\usepackage{hyperref}
\usepackage{lscape}
\usepackage{abstract}
\usepackage{authblk}

\allowdisplaybreaks

\AtBeginEnvironment{tabular}{\footnotesize}
\AtBeginEnvironment{tabularx}{\footnotesize}
\AtBeginEnvironment{longtable}{\footnotesize}

\titleformat*{\section}{\normalfont\fontsize{16}{19}\bfseries}
\titleformat*{\subsection}{\normalfont\fontsize{14}{17}\itshape}
\titleformat*{\subsubsection}{\normalfont\fontsize{14}{17}\selectfont}

\begin{document}

\title{Bootstrapping Two-Nucleon Effective Field Theories}

\author[a]{\normalsize Q.N. Micha-Mba\thanks{quiricomba@usal.es}}
\author[a]{M.S. Sánchez\thanks{mariosanchez2@usal.es}}
\author[a,b]{P.G. Ortega\thanks{pgortega@usal.es}}
\author[c]{J.A. Oller\thanks{oller@um.es}}
\author[a,b]{D.R. Entem\thanks{entem@usal.es}}
\affil[a]{\it Departamento de Física Fundamental, 
Universidad de Salamanca, E-37008 Salamanca, Spain}
\affil[b]{\it Instituto Universitario de Física Fundamental y Matemáticas (IUFFyM),\protect\\
Universidad de Salamanca, E-37008 Salamanca, Spain}
\affil[c]{\it Departamento de Física, 
Universidad de Murcia, E-30071 Murcia, Spain}

\maketitle

\begin{abstract}
Chiral EFT yields singular potentials that require regularization and renormalization when implemented in a dynamical equation such as the Lippmann--Schwinger equation. We employ two different approaches, renormalization with contact terms---as is most commonly done in chiral EFT---and the exact N/D method with multiple subtractions. We start with a toy model in which we can control the finite-range expansion of the potential, treating the full potential as the `exact' theory. To assess the statistical consistency of the approaches with the full theory, we use the bootstrap technique. We apply the same framework to study the consistency of chiral EFT at LO and NLO with the Granada phase-shift analysis in the $^1S_0$ two-nucleon partial wave. Our results show that the NLO potential significantly extends the energy range over which the theory remains valid. 
\end{abstract}

\bigskip
\section{Introduction}

For the last few decades, the effective field theory (EFT) framework has been widely used to study the two-nucleon ($NN$) problem. The original idea was proposed by Weinberg~\cite{Weinberg:1990rz,Weinberg:1991um}, who noticed that chiral perturbation theory ($\chi$PT) by itself fails to describe the $NN$ system. The reason is that iterative {(or reducible)} diagrams break the chiral expansion, as they are enhanced by \textit{{negative}} powers of the small ratio between the low-momentum scale $Q$ (typical magnitude of the external three-momenta), of the same order as the pion mass $m_\pi$, and the high-momentum scale $M_{\text{hi}}$, approximately set by the nucleon mass $m_N$ and the chiral-symmetry breaking scale. This idea is supported by the unnaturally large $^1S_0$ and $^3S_1$ scattering lengths, {which, respectively, give rise to a virtual state and a shallow bound state (the deuteron)}, since these phenomena attest to important non-perturbative effects. Weinberg thus postulated to calculate the $NN$ potential with the rules of $\chi$PT, and then include it in an \textit{exact} Lippmann--Schwinger (LS) or Schr\"odinger equation where an infinite number of reducible diagrams is taken into account---a program originally followed in Refs. \cite{ORDONEZ1992459,PhysRevLett.72.1982,PhysRevC.53.2086}.

\medskip
Loop diagrams generate infinities that must cancel out with some given constants of the theory. In the perturbative case, the infinities are polynomials in the ingoing and outgoing $NN$ momenta that can be absorbed by derivative four-nucleon Lagrangian terms encoding the so-called contact terms (or counterterms). Weinberg proposed a power counting where naive dimensional analysis (or `naturalness') dictates which Lagrangian terms are to be included at a certain order of the effective expansion, together with the diagrams that have to be considered at the same order. However, it is under debate whether or not the same
contact terms that renormalize the perturbative theory also suffice to renormalize the non-perturbative LS equation. The issue of renormalization 
in nuclear EFT has been profusely discussed in the literature---see, for instance, Refs.~\cite{Kaplan:1996xu,Kaplan:1998tg,Kaplan:1998we,PhysRevC.72.054006,Birse:2005um,Valderrama:2009ei,PavonValderrama:2011fcz,Long:2011qx,Long:2011xw,Long:2012ve,Epelbaum:2006pt,Epelbaum:2009sd,Machleidt:2010kb,Marji:2013uia,Epelbaum:2020maf,Epelbaum:2017byx,Epelbaum:2017tzp,Epelbaum:2018zli,Epelbaum:2019msl,vanKolck:2020llt,Peng:2024aiz,Yang:2021vxa,Baru:2019ndr,Gasparyan:2022isg,Gasparyan:2021edy}.

\medskip
{Since the EFT amounts to expansions of observables in (positive) powers of $Q/M_{\text{hi}}$,} it is intended to work better for small energies. We thus present an expansion of the potential starting from the longer-range interactions, while higher-order terms give shorter-range contributions. The renormalization process mimics the unresolved short-range physics through zero-range interactions given by contact terms.

\medskip
In the Weinberg proposal, the multi-pion exchange potentials coming from $\chi$PT are singular. This implies that they cannot be included in the LS equation directly, but require regularization and renormalization. Regularization is usually performed through a regulator function that depends on an ultraviolet cutoff $\Lambda$, and then renormalization is performed by fitting the values of the contact terms to the scattering data. Here, two different points of view arise depending on the values for the cutoff $\Lambda$ that are deemed reasonable.

\medskip
From the standpoint of quantum field theory, one would like to completely remove the cutoff by taking $\Lambda\to\infty$. However, this 
{is problematic in a fully non-perturbative framework}
because all the contact terms become irrelevant operators, except the lowest-order one in the singular attractive
case~\cite{Case:1950an,Frank:1971xx,PavonValderrama:2005gu,PavonValderrama:2005wv,PavonValderrama:2005uj}. {This element might explain the issues of this approach in the description of scattering phase shifts and also in its difficulties to yield a convergent pattern at higher orders~\cite{Zeoli:2012bi}, unless those are treated in perturbation theory (see Ref. \cite{vanKolck:2020llt} for a review).} 

\medskip
The other point of view is to keep the cutoff value somewhere between the soft scale $Q\sim100\,\text{MeV}$ and
the hard scale $M_{\text{hi}}\sim1\,\text{GeV}$, that is, in a window of values
around 500 MeV. This picture has been followed by potential models with great success~\cite{Machleidt:2011zz,Epelbaum:2008ga}. 
Following this idea, high-precision $NN$
potentials have been constructed at N$^3$LO and higher orders. But, even though contact terms allow one to obtain a mild $\Lambda$ dependence {for such restricted cutoff window}, small cutoff artifacts appear.

\medskip
An alternative to the infinite-cutoff limit has recently been proposed, {based on using the old N/D method{\cite{Chew:1960iv}}}---an idea that was first applied to pion scattering and has been primarily employed in perturbative frameworks. In this new proposal, referred to as the exact N/D method~\cite{Entem:2016ipb,Oller:2018zts}, the discontinuity of the potential along the left-hand cut (LHC) is included in an integral equation (IE) that contains the infinite sum of iterative diagrams. The solution of this IE---the exact LHC discontinuity of the amplitude---is the input of the N/D method and does not need any regularization procedure. It has been shown to be equivalent to the LS equation for regular interactions and{---when the solution exists---}to renormalization with one contact term in the infinite-cutoff limit (equivalent to renormalization with boundary conditions ~\cite{Entem:2021kvs}). However, as we will show, this strategy can go beyond the LS (or Schr\"odinger) formulation. Also, it is interesting to notice that the N/D method with multiple
subtractions is equivalent to solving the LS equation with exact (not regularized) contact terms using dimensional-regularization (DR) techniques
when the finite-range potential is regular~\cite{Entem:2025siq}.

\medskip
In all pictures there are unknown parameters that mimic the short-range part of the interaction and have to be obtained from the scattering data. In order to see the consistency of the data and the theory, we perform a bootstrap calculation with pseudo-experimental data. We will first apply this idea to a toy model and then use it for the $^1S_0$ $NN$ partial wave.

\medskip
This paper is organized as follows. In Section~\ref{Sct} we explain the renormalization procedure with contact terms. In Section~\ref{ND}
we introduce the exact N/D method. In Section~\ref{TM} we present the toy model we want to analyze. In Section~\ref{fits} we introduce
the bootstrap technique, apply it to our toy model, and give the results. In Section~\ref{chiEFT} we apply the technique to the $^1S_0$ $NN$ partial wave.
We end in Section~\ref{conc} with some conclusions.

\section{Renormalization with contact terms}
\label{Sct}

The scattering problem is usually solved with the LS equation, which reads for the uncoupled case 
\begin{eqnarray}
T(p',p;k) &=& V(p',p) + \,\frac{m_N}{2\pi^2} \int_0^\infty dq\,\frac{V(p',q)\,q^2\,T(q,p;k)}{k^2-q^2+i\epsilon}\,.
\end{eqnarray}
However, if the potential is singular, we cannot include it directly in the LS equation. In turn, we will regularize the potential and renormalize the solution with contact terms.
Here we will use the {`super-Gaussian', non-local} regulator function
\begin{eqnarray}
	f_\Lambda(p',p) &=& e^{-{({p'}^6+p^6)}/{\Lambda^6}}
\end{eqnarray}
with $\Lambda=600$ MeV, and take the potential
\begin{eqnarray}
	\widetilde V(p',p) &=& f_\Lambda(p',p) V(p',p)\,, 
\end{eqnarray}
where $V(p',p)$ is the unregularized interaction. We renormalize the result, partially removing the effect of the regulator, by including contact terms {that for an $S$-wave read up to ${\mathcal{O}(Q^4)}$} 
\begin{eqnarray}
	\widetilde V_{\text{ct}}(p',p) &=& f_\Lambda(p',p) \left[C_0 + C_2 (p'^2+p^2)+C_4 (p'^4+p^4) + \widetilde{C}_4 p'^2p^2+\ldots\right]\,.
\end{eqnarray}
The values of the contact terms $C_0,\,C_2,\,\ldots$ depend on the cutoff value $\Lambda$ and the shape of the function $f_\Lambda(p',p)$, where the latter has been chosen so that the contact terms will not change at first order in $1/\Lambda$. These values will be obtained by fitting the phase shifts as explained below. The effect of $C_4$ and $\widetilde{C}_4$ is very similar \cite{Beane:2000fi}, and we will only include one of them.
In fact, it has been shown \cite{Entem:2025siq} that, in DR, their contribution to the on-shell amplitude is exactly the same.

\section{The exact N/D method}
\label{ND}

The N/D method was introduced long ago by Chew and Mandelstam~\cite{Chew:1960iv}. 
The idea is to separate the cuts of the on-shell amplitude $T(A)\equiv T(k,k;k)$, where
$A\equiv k^2$. Apart from the aforementioned LHC due to pion exchanges, there is a unitarity or right-hand cut (RHC) due to resummation of intermediate $NN$ states. One thus writes
\begin{eqnarray}
	T(A) &=& \frac{N(A)}{D(A)}\,,
\end{eqnarray}
where $N(A)$ has only an LHC, and $D(A)$ has only an RHC. On the RHC,
\begin{eqnarray}
	{\rm Im}D(A) &=& - \rho(A) N(A)\,,
\end{eqnarray}
where $\rho(A)$ is the phase-space factor. On the LHC,
\begin{eqnarray}
	{\rm Im}N(A) &=& D(A) \Delta T(A)\,,
\end{eqnarray}
with $\Delta T(A)={\rm Im}T(A)$. Then, subtracted dispersion relations yield the IEs
\begin{eqnarray}
	D(A) &=& 1 - \frac A \pi \int_0^\infty d\omega_R \frac{\rho(\omega_R) N(\omega_R)}{(\omega_R-A)\omega_R}\,;
	\nonumber \\
	N(A) &=& \frac 1 \pi \int_{-\infty}^L d\omega_L \frac{D(\omega_L)\Delta T(\omega_L)}{\omega_L-A}\,,
\end{eqnarray}
where $L$ sets the onset of the LHC. 
Since $T(A)$ remains unchanged when multiplying $N(A)$ and $D(A)$ by the same factor, we need to fix at least one of them at a certain value of $A$, so we set the threshold condition $D(0)=1$ with one single subtraction on $D$. For regular interactions, this is equivalent to solving the LS equation. We will call this the N/D$_{01}$ solution.

\medskip
However, more subtractions
can be made, and we can fix the value of the {$S$-wave effective-range expansion (ERE)} parameters
\begin{eqnarray}
	k \cot \delta &=& -\frac{1}{a} + \frac r 2 k^2 + \sum_{i=2}^\infty v_i k^{2i}\,.
\end{eqnarray}
These solutions will be called N/D$_{nd}$, where we make $n$ subtractions on $N$ and $d$ subtractions on $D$ by fixing the first ($n+d-1$) ERE parameters.

\medskip
The N/D method requires the input of $\Delta T(A)$ along the LHC. Traditionally, such function was evaluated in perturbation theory, and this was the reason why the N/D method was not equivalent to the non-perturbative LS equation. In turn, the recently developed exact N/D method~\cite{Entem:2016ipb,Oller:2018zts}
introduced the IE {for the uncoupled case}
\begin{eqnarray}
        \Delta \widehat T(\nu,\bar k) &=& \Delta \widehat V(\nu,\bar k) + \theta(\bar k-m_\pi)
        \theta(\bar k-2m_\pi-\nu) \frac{m_N}{2\pi^2}
        \int_{\nu+m_\pi}^{\bar k-m_\pi}\!d\nu_1 \frac{\Delta \widehat V(\nu,\nu_1)\,\nu_1^2S(\nu_1)\,\Delta \widehat T(\nu_1,\bar k)}{\bar k^2-\nu_1^2}\,,
        \label{IntEq}
\end{eqnarray}
with 
$\bar k = -ik$, and
\begin{eqnarray}
        S(\nu_1) &=& \frac{1}{(\nu_1+i\epsilon)^{2l+2}} + \frac{1}{(\nu_1-i\epsilon)^{2l+2}}\,;
        \\
        2\Delta V(\nu_1,\nu_2) &=&
        \frac{2\Delta \widehat V(\nu_1,\nu_2)}{\nu_1^{l+1} \nu_2^{l+1}}
        \nonumber \\
        &=& \lim_{\epsilon\to 0} \lim_{\delta\to 0}
        \big[
        {\rm Im} V(i\nu_1+\epsilon-\delta,i\nu_2+\epsilon)-{\rm Im} V(i\nu_1+\epsilon+\delta,i\nu_2+\epsilon)
        \big]\,;
        \\
        2\Delta T(\nu,\bar k) &=&
        \frac{2\Delta \widehat T(\nu,\bar k)}{\nu^{l+1} \bar k^{l+1}}
        \nonumber \\
        &=& \lim_{\epsilon\to 0} \lim_{\delta\to 0}
        \big[
        {\rm Im} T(i\nu+\epsilon-\delta,i\bar k+\epsilon)-{\rm Im} T(i\nu+\epsilon+\delta,i\bar k+\epsilon)
        \big]\,,
\end{eqnarray}
where the LHC discontinuity of the on-shell amplitude is
\begin{eqnarray}
        \Delta T(A) &=& -\frac{\Delta \widehat T(-\bar k,\bar k)}{\bar k^{2l+2}}\,.
\end{eqnarray}
Notice that Eq. (\ref{IntEq}) is always finite and no cutoff is introduced---an element that allows us to apply this approach to singular interactions as well without including any regulator function.

\section{A toy model}
\label{TM}

Following the idea proposed by Epelbaum {\it et al.}~\cite{Epelbaum_2018} to show the limitations of infinite-cutoff renormalization
procedures, we consider the toy model~\cite{Entem:2021kvs}
\begin{eqnarray}
        V(r) &=& V_1(r) +  V_2(r) +  V_3(r) +  V_4(r)\,, \\
        \label{potTM}
        V_1(r) &=& -\alpha \frac{e^{-m_\pi r}}{r}\,;
        \\
        V_2(r) &=& \alpha_1 \frac{e^{-2m_\pi r}}{r^3}\,;
        \\
        V_3(r) &=& -\alpha_1 (m_2-2m_\pi) \frac{e^{-m_1r}}{r^2}\,;
        \\
        V_4(r) &=& -\alpha_1 \frac{e^{-m_2r}}{r^3}\,.
\end{eqnarray}
It has a long-range regular part given by $V_1(r)$, similar to the contribution of the one-pion exchange (OPE) interaction in the $^1S_0$ 
partial wave. Then three singular terms $V_2(r)$, $V_3(r)$, and $V_4(r)$ are added. They combine so that
\begin{eqnarray}
	V(r) \xrightarrow[r\to 0]{} \frac{-\alpha +2\alpha_1(m_\pi-\frac{m_2}{2})(m_\pi+\frac{m_2}{2}-m_1)}{r}\,,
\end{eqnarray}
so the full potential is regular. The inverse of the range of $V_2(r)$ is $2m_\pi$, simulating singular two-pion exchange (TPE) interactions; the inverses of the ranges of $V_3(r)$ and $V_4(r)$, respectively $m_1$ and $m_2$, are both of the order of 1 GeV, simulating the unknown short-range components of the theory. The problem we want to address is whether knowing only the long-range part of the interaction---in this case $V_1(r)$ and, eventually, $V_2(r)$---may suffice to reproduce in detail the results of the full potential $V(r)$ using renormalization techniques. The sign of $\alpha_1$ sets the nature of $V_2(r)$---either singular repulsive ($\alpha_1>0$) or singular attractive ($\alpha_1<0$). This distinction has important consequences for the case where one intends to renormalize with one single renormalization condition.

\medskip
Although this is not a field theory, we will refer to results considering $V(r)$ as the `full theory', including only $V_1(r)$ as `leading order' (LO), and including $V_1(r)+V_2(r)$ as `next-to-leading order' (NLO). This is of course not a power counting---in fact, we are not fixing the number of renormalization
conditions ({determined by} the number of contact terms or subtractions in the N/D method) at each order. Our purpose is rather to test the results considering different possibilities. Here, `LO' and `NLO' should be understood as the orders of the finite-range interactions, from longer to shorter ranges.

\begin{table}
\caption{Parameters used for the toy model. For $\alpha_1$ we use two different values to simulate
the singular repulsive ($\alpha_1>0$) and singular attractive ($\alpha_1<0$) cases.\label{param}}
\centering
\begin{tabular}{ccc}
\noalign{\smallskip}\hline
$m_\pi$ & 138.5 & MeV  \\
\noalign{\smallskip}\hline\noalign{\smallskip}
$m_1$ & 1000 & MeV  \\
\noalign{\smallskip}\hline\noalign{\smallskip}
$m_2$ & 1200 & MeV  \\
\noalign{\smallskip}\hline\noalign{\smallskip}
$m_N$ & 938.919 & MeV \\
\noalign{\smallskip}\hline\noalign{\smallskip}
$\alpha$ & 0.1 & \\
\noalign{\smallskip}\hline\noalign{\smallskip}
$\alpha_1$ & $+5.0$ & GeV$^{-2}$ \\
           &$-3.0$ & GeV$^{-2}$ \\
\noalign{\smallskip}\hline
\end{tabular}
\end{table}

\begin{figure*}
\centering
\includegraphics[width=7 cm]{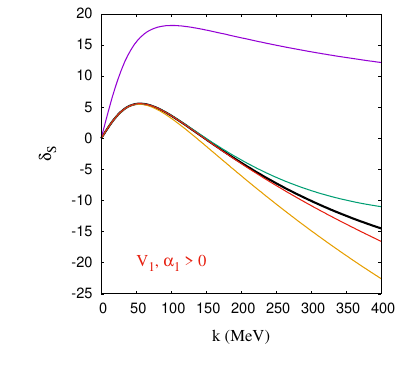}
\includegraphics[width=7 cm]{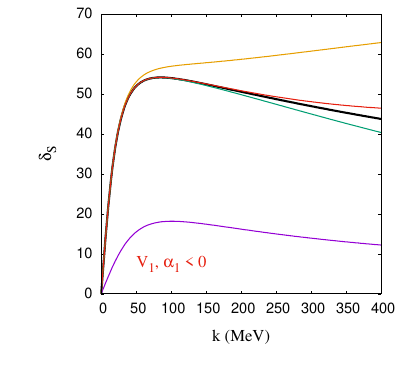}
\includegraphics[width=7 cm]{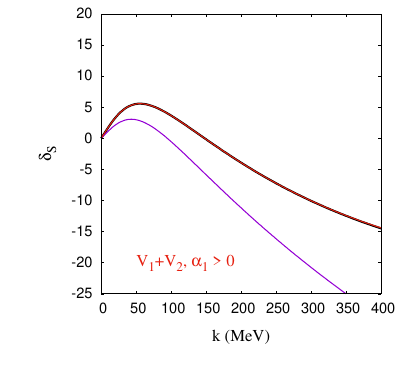}
\includegraphics[width=7 cm]{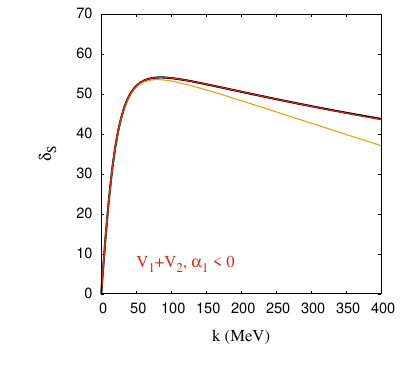}
\includegraphics[width=7 cm]{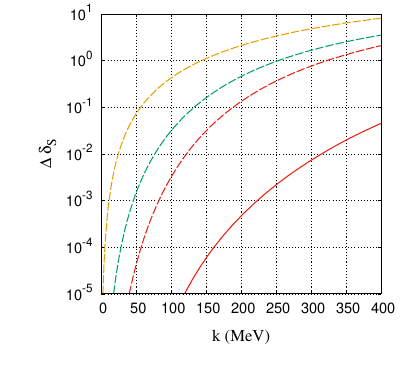}
\includegraphics[width=7 cm]{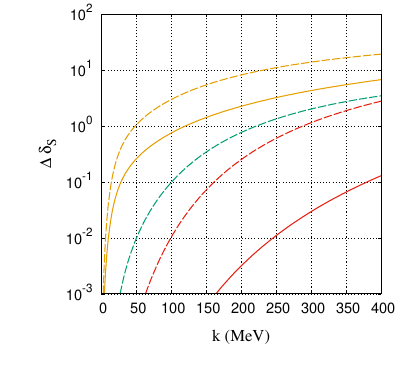}
\caption{\label{figteo} Panels on the left are for the singular repulsive case while on the right for the singular attractive case. In all panels the black line corresponds
to the result of the full theory. Other lines correspond to the result of the N/D method at LO (upper panels) and NLO (middle panels).
The lower panels show the difference between the full theory and the result at LO (dashed lines) and NLO (solid lines).
The color code for the lines is purple for N/D$_{01}$, gold for N/D$_{11}$, green for N/D$_{12}$, and red for N/D$_{22}$.} 
\end{figure*}

\medskip
The potential in momentum space is given by
\begin{eqnarray} \label{potv1}
        V_1(q) &=& \frac{4\pi\alpha}{q^2+m_\pi^2}\,;
        \\
        \label{potv2}
        V_2(q) &=& 8 \pi \alpha_1 m_\pi \left[ A_\pi(q) + \frac{ L_\pi(q)}{2m_\pi} \right]\,;
        \\
        V_3(q) &=& 4\pi\alpha_1 (m_2-2m_\pi) A_1(q)\,;
        \\
        V_4(q) &=& -4 \pi \alpha_1 m_2 \left[ A_2(q) + \frac{L_2(q)}{m_2} \right]\,,
\end{eqnarray}
where 
\begin{eqnarray}
        L_\pi(q) &=& \log\bigg(\frac{\sqrt{q^2+4m_\pi^2}}{2m_\pi}\bigg)\,;
        \\
        A_\pi(q) &=& \frac{1}{q} \arctan\bigg(\frac{q}{2m_\pi}\bigg)\,;
        \\
        L_i(q) &=& \log\bigg(\frac{\sqrt{q^2+m_i^2}}{m_i}\bigg)\,;
        \\
        A_i(q) &=& \frac{1}{q} \arctan\bigg(\frac{q}{m_i}\bigg)\,.
\end{eqnarray}
Note that for natural values of $\alpha$ and $\alpha_1$, that is, $\alpha={\cal O}(1)$ and $\alpha_1={\cal O}(M_{\text{hi}}^{-2})$, one finds $V_2(Q)/V_1(Q)={\cal O}(Q^2/M_{\text{hi}}^{2})$ in the chiral counting with $m_\pi\sim Q$. This follows from the explicit expressions given for $V_1$, $V_2$, $L_\pi$, and $A_\pi$.

\medskip
In Table~\ref{param} we give the parameters of the model we are using. In Fig.~\ref{figteo} we compare {at the level of phase shifts} several results from the N/D method with the full-theory result. 
The left (right) panels show the singular repulsive (attractive) case. The upper (middle) panels consider the potential at LO (NLO); deviations from the full theory are given by the lower panels. 

\medskip
Since $V_1$ is regular, solutions are well-defined with none (N/D$_{01}$), one (N/D$_{11}$), two (N/D$_{12}$), and three (N/D$_{22}$) renormalization conditions. (We will not impose more than three renormalization conditions in this work.) Both for $\alpha_1>0$ and $\alpha_1<0$, we can reproduce the exact phase shifts to higher energies by increasing the number of subtractions, as expected. 

\medskip
The potential becomes singular when we include $V_2$, and it is not always possible to impose a certain number of renormalization conditions. In particular: 
\begin{itemize}
    \item[$\bullet$] For the singular repulsive case, we cannot impose one single renormalization condition; instead, we have the regular solution N/D$_{01}$. As shown in Fig.~\ref{figteo}, the unsubtracted NLO result is worse than the once-subtracted LO one---there is no convergent pattern when we include more physics through $V_2$, which is one of the limitations mentioned above. 
    
    \item[$\bullet$] The situation is the opposite for the singular attractive case, where the regular solution is not well-defined, but we can fix the scattering length $a$ (N/D$_{11}$ solution), and the once-subtracted NLO result indeed improves the once-subtracted LO one, as shown in Fig.~\ref{figteo}.  
\end{itemize}

However, {contrary to what happens with the infinite-cutoff renormalization of the LS equation,} there are further N/D solutions. While N/D$_{12}$ does not converge in either case (so it is not shown), N/D$_{22}$ does converge in both cases---as one checks by fixing, apart from $a$, the effective range $r$ and the shape parameter $v_2$. In both the repulsive and attractive cases, there is very good agreement with the exact solution up to the limit we consider ($k=400$ MeV). It is also clear that the NLO description is better than the LO one. (Notice that the N/D$_{22}$ solution at LO is accurate up to $k\approx 200$ MeV for both cases, and then starts to deviate.)

\begin{table}
\caption{\label{EREV} ERE parameters of the full potential $V(r)$.}
\centering
\begin{tabular}{ccl}
\noalign{\smallskip}\hline
$a$ & $-$0.61518 & fm   \\
\noalign{\smallskip}\hline\noalign{\smallskip}
$r$ & 28.148 & fm  \\
\noalign{\smallskip}\hline\noalign{\smallskip}
$v_2$ & 19.007 & fm$^3$  \\
\noalign{\smallskip}\hline
\end{tabular}
\end{table}

\medskip
It is important to note that in the previous results the exact values of $a$, $r$, and $v_2$ are known and given by the full potential $V$, see Table~\ref{EREV}. {(Note that the N/D method allows us to very accurately compute an arbitrarily high number of ERE parameters in virtue of the strategy introduced in Ref. \cite{Oller:2014uxa}.)} In the real world, this does not happen, and the ERE parameters need to be obtained from the experimental information given by the scattering data. Of course, its determination will depend on the precision of the data, so an important question arises:  \textit{which precision is needed} to resolve the details of the interaction generated by $V_2$, since considering only $V_1$ may already suffice to reproduce low-precision data by mimicking the effect of $V_2$ on the values of the unknown parameters.

\medskip
The expected precision of the theory at different orders and different renormalization conditions is given by the lower panels in Fig.~\ref{figteo}, where we display the difference between the several approaches considered and the exact result. For example, if we demand a precision of $0.01^\circ$ in the repulsive case, the lower left panel shows that, provided three renormalization conditions, the LO (NLO) theory is accurate enough up to $\approx 150$ MeV ($\approx 300$ MeV).

\section{Fits}
\label{fits}

We are going to analyze how we can extract short-range information from the scattering data. For that we propose a theoretical experiment where, for simplicity, our data will be directly the phase shifts. We will generate experiments from our known full theory considering random data points following a Gaussian distribution with a mean value given by the exact result and a mean deviation given by a certain error $\Delta \delta$. We will take the on-shell momentum $k$ ranging from 1 MeV to 400 MeV equally spaced in 1 MeV. 

\medskip
Since we are working with a theoretical experiment, it is important to notice that many problems that appear in the real case are not present. For example, the error is fixed by us, so problems of underestimating or overestimating the experimental error are not present. In addition, there are no outliers due to uncontrolled systematic errors in the experiment. Thus, the usual problem faced in the context of $NN$ phenomenology---either accepting or rejecting data groups---is not present by construction, since all data are built consistently.

\medskip
We will perform a $\chi^2$ fit to our pseudo-experimental data up to $k_{\rm max}$.
Since we are using randomly distributed data, we should expect $\chi^2$ values with a probability distribution given by
\begin{eqnarray}
	f(\chi^2) &=& \frac{1}{2^{\nu/2}\Gamma(\nu/2)} e^{-\frac{\chi^2}{2}} (\chi^2)^{\nu/2-1}\,,
	\label{chi2dis}
\end{eqnarray}
with $\nu$ the number of degrees of freedom given by the number of data minus the number of parameters of the model. This means that, given a certain value $\chi^2_0$, the probability of an experiment with a higher $\chi^2$ value is given by
\begin{eqnarray}
	P(\chi^2 > \chi^2_0) &=& \int_{\chi^2_0}^\infty f(\chi^2) d \chi^2\,,
\end{eqnarray}
and, of course, the probability of getting a lower value is given by $P(\chi^2 < \chi^2_0)=1-P(\chi^2 > \chi^2_0)$. We define a certain
confidence level $\alpha$ such that there is a probability $\alpha$ of obtaining a value in the range $(\chi^2_{\rm min},\chi^2_{\rm max})$,
$\frac{1-\alpha}{2}$ of getting a value lower than $\chi^2_{\rm min}$ (the $\frac{1-\alpha}{2}$ quantile), and $\frac{1-\alpha}{2}$ of getting a value bigger than
$\chi^2_{\rm max}$ (the $1-\frac{1-\alpha}{2}=\frac{1+\alpha}{2}$ quantile). The range is thus defined by the equations
\begin{eqnarray}
	\int_{\chi^2_{\rm min}}^\infty f(\chi^2) d \chi^2 &=& \frac{1+\alpha}{2}\,;
	\\
	\int_{\chi^2_{\rm max}}^\infty f(\chi^2) d \chi^2 &=& \frac{1-\alpha}{2}\,.
\end{eqnarray}
Notice that the range $(\chi^2_{\rm min},\chi^2_{\rm max})$ depends on the number of degrees of freedom $\nu$. For large values of $\nu$,
\begin{eqnarray}
	\chi^2_{\rm max}/{\rm dof}-\chi^2_{\rm min}/{\rm dof} &\approx& 2\sqrt{{2}/{\nu}}\,.
\end{eqnarray}
If we consider a value of $\alpha$ close to 1, obtaining values outside the confidence range means that, either we obtained a very unlikely value, or the theory we use to describe the data is not good enough for the precision given by the data. We will consider the second case, and this will help us to discriminate when our theory is consistent with the data. Notice that the accuracy of the data is important, since less accuracy in the data means that less accuracy is also needed in the theory.

\medskip
If we only use one experiment (one set of data), then $\chi^2$ tells us the probability of obtaining such data given a certain theory. However, what we want to quantify is the probability of having a theory given some data. To this end, we could use a Bayesian approach, which is the common way of doing things nowadays {(see e.g. Refs. \cite{Furnstahl:2015rha, Melendez:2017phj, Melendez:2019izc}).
In this framework, the coefficients of the EFT expansion and its truncation error are treated as random variables constrained by prior information that encodes standard EFT assumptions, such as naturalness and a convergent power counting. The information contained in the calculated orders is then used to update these priors and infer posterior probability distributions for observables and their associated uncertainties. This provides a statistically consistent way to estimate truncation errors, assign confidence levels, and analyze convergence patterns, instead of giving purely heuristic error estimates.} 

\medskip
However, in our theoretical experiment we will perform a bootstrap analysis. 
The bootstrap technique
was introduced by Efron~\cite{10.1214/aos/1176344552} as a method to obtain distributions from data sets.
Essentially, this was the idea employed by the authors of Ref.~\cite{NavarroPerez:2014bca}, where a bootstrap study was applied to the OPE potential supplemented by short-range interactions in the form of delta-shell terms in coordinate space.
So we will generate many sets of data and fit all of them with a given theory to examine the statistical consistency of the fits. Therefore, the parameters of the short-range interaction are treated as random variables, and its probability distribution function is obtained from bootstrap.

\medskip
For the experiment to be statistically consistent, not only should a good $\chi^2$ distribution (\ref{chi2dis}) be obtained. The $\chi^2$ is
the sum of the square of the residuals, namely
\begin{eqnarray}
\chi^2 &=& \sum_{i=1}^{N}\left(\frac{y_i - y(x_i)}{\delta y_i}\right)^2\,, 
\end{eqnarray}
where $N$ is the number of data, $y_i$ are the data with errors $\delta y_i$, and $y(x_i)$ is the model (theory) we want to test. If the fit is consistent, then the residuals 
should follow a normal distribution function given by
\begin{eqnarray}
	{\mathcal N}(\mu,\sigma) &=& \frac{1}{\sqrt{2\pi}\sigma} e^{-\frac{(x-\mu)^2}{2\sigma^2}}\,,
\end{eqnarray}
with $\mu=0$ and $\sigma=1$---this is an important statistical test of the $\chi^2$ fit. 
However, sometimes even with normally distributed residuals we can see statistical deviations in the data. So we will also check the difference between the mean of the fits and the exact value of the theory.

\medskip
For definiteness, in what follows we will always consider the singular repulsive ($\alpha_1>0$) case, since the singular attractive ($\alpha_1<0$) one is conceptually very similar for what concerns our bootstrap analysis.

\subsection{Full theory}

The exact theory is given by the N/D$_{01}$ solution due to the full potential $V(r)$. However, since the potential is regular, the N/D$_{11}$, N/D$_{12}$, N/D$_{22}$, \ldots \,solutions will also be well-defined. Note that all these solutions, which are obtained by fixing the first $p$ ERE parameters with $p=1,2,3,\ldots\,$, include the exact theory when such ERE parameters are adjusted to the values predicted by the N/D$_{01}$ solution. Hence, if we implement the full potential in the N/D method, the result should be accurate regardless of the precision, the number of subtractions, or the momentum range of the data. This element will help us to check the statistical consistency of our calculations.

\begin{figure*}[th]
\centering
\includegraphics[width=5 cm]{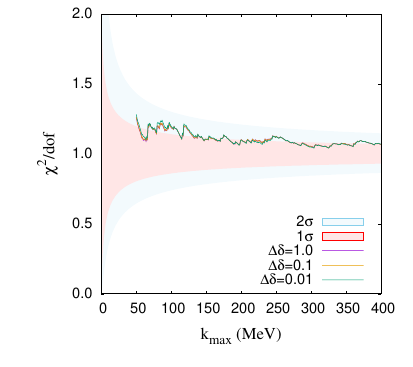}
\includegraphics[width=5 cm]{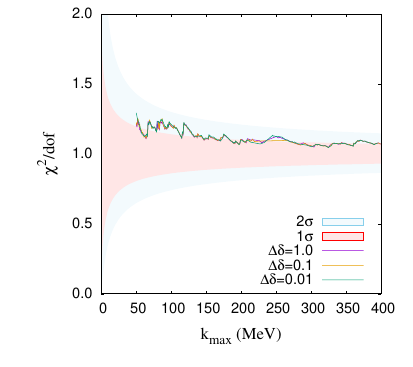}
\includegraphics[width=5 cm]{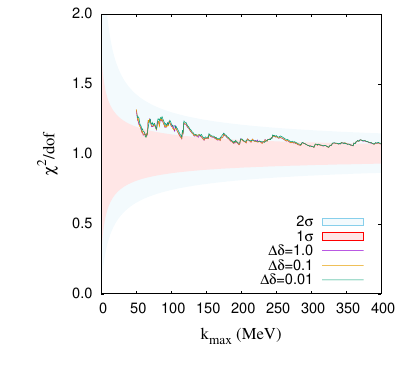}
\caption{\label{figV4} $\chi^2$/dof for fits with data between 1 MeV and $k_{\rm max}$. The red (blue) band shows the 1$\sigma$ (2$\sigma$) confidence level. Data are generated as explained in the text with $\Delta \delta = 1^\circ$ (purple line), $0.1^\circ$ (gold line) and $0.01^\circ$ (green line). In all cases the full theory is considered. From left
to right, panels correspond to the N/D$_{11}$, N/D$_{12}$, and N/D$_{22}$ solutions.}
\end{figure*}

\medskip
First, we make a $\chi^2$ fit to one single experiment. We consider a certain phase-shift uncertainty $\Delta \delta$ for all energies. In Fig.~\ref{figV4} we show the results for the N/D method including the full theory. We plot $\chi^2/$dof for fittings up to a certain $k_{\rm max}$, displaying the 1$\sigma$ ($\alpha=0.6827$) and 2$\sigma$ ($\alpha=0.9545$) bands in the red and blue areas, respectively. Notice that the number of degrees of freedom is the number of data, given by $k_{\rm max}/$(1 MeV), minus the number of parameters in the fit. This means that if we fit our experimental data with the correct theory we have a 68$\%$ chance to get a $\chi^2$ value in the red area, a 95$\%$ in the blue area (which subsumes the red one), and a $<5\%$ chance to be outside, since those are the probabilities to generate the data from the correct theory. From left to right, we give the results corresponding to the N/D$_{11}$, N/D$_{12}$, and N/D$_{22}$ solutions. We consider three different phase-shift errors---$\Delta \delta = 1.0^\circ,\,0.1^\circ,\,\text{and}\,0.01^\circ$. In all cases, $\chi^2$ is clearly inside the 2$\sigma$ band, which shows that all theories---regardless of the number of renormalization conditions, energy region, and precision---are expected to explain the data.

\begin{table*}[th]
\caption{\label{bootchi2} Bootstrap to 2000 experiments: $\chi^2/$dof from the bootstrap (fourth column) and from the theoretical
distribution (\ref{chi2dis}) (fifth column).}
\centering
\begin{tabular}{ccccc}
\noalign{\smallskip}\hline
$k_{\rm max}$(MeV) & N/D & $\Delta \delta$ & $\chi^2/$dof  & $\chi^2/$dof (th.)  \\
\noalign{\smallskip}\hline\noalign{\smallskip}
400 & N/D$_{11}$ & 0.01$^\circ$ & $1.000 \pm 0.071$ & $0.998^{+0.072}_{-0.069}$ \\
\noalign{\smallskip}\hline\noalign{\smallskip}
400 & N/D$_{12}$ & 0.01$^\circ$ & $1.001 \pm 0.070$ & $0.998^{+0.072}_{-0.069}$ \\
\noalign{\smallskip}\hline\noalign{\smallskip}
400 & N/D$_{22}$ & 0.01$^\circ$ & $0.999 \pm 0.071$ & $0.998^{+0.073}_{-0.069}$ \\
\noalign{\smallskip}\hline\noalign{\smallskip}
400 & N/D$_{22}$ & 0.1$^\circ$ & $0.997 \pm 0.071$ & $0.998^{+0.073}_{-0.069}$ \\
\noalign{\smallskip}\hline\noalign{\smallskip}
300 & N/D$_{22}$ & 0.01$^\circ$ & $1.000 \pm 0.082$ & $0.998^{+0.084}_{-0.080}$ \\
\noalign{\smallskip}\hline\noalign{\smallskip}
200 & N/D$_{22}$ & 0.01$^\circ$ & $1.000 \pm 0.099$ & $0.997^{+0.104}_{-0.097}$ \\
\hline\noalign{\smallskip}
\end{tabular}
\end{table*}

\begin{figure}[th]
\centering
\includegraphics[width=5 cm]{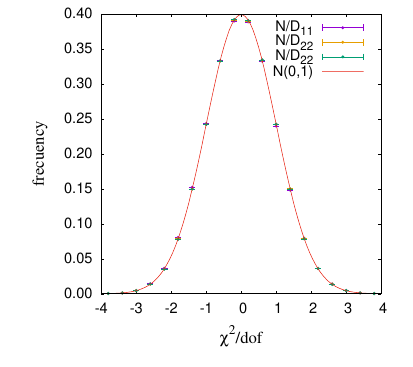}
\caption{\label{resV4} Residuals for $\Delta \delta=0.01^\circ$ in three cases. 
The purple, gold, and green dots correspond to N/D$_{11}$ with $k_{\rm max}=400$ MeV, N/D$_{22}$ with $k_{\rm max}=400$ MeV, and N/D$_{22}$ with $k_{\rm max}=200$ MeV, respectively.
The solid line corresponds to the ${\mathcal N}(0,1)$ distribution.
}
\end{figure}

\medskip
Now we do a bootstrap by considering 2000 experiments and check the statistical consistency. In Table~\ref{bootchi2} we give in the fourth column the mean $\chi^2$/dof for all experiments with their mean deviations at the 1$\sigma$ level. This is done for different N/D solutions, $k_{\rm max}$, and $\Delta \delta$.
The fifth column displays the results of the theoretical distribution (\ref{chi2dis}). Notice that for the theoretical
distribution the central value corresponds to the value
that has 50$\%$ probability to get higher and lower values, 
and the limits are the 1$\sigma$ ones defined previously, which explains the small asymmetry. The values show good agreement.

\begin{figure*}[th]
\centering
\includegraphics[width=5 cm]{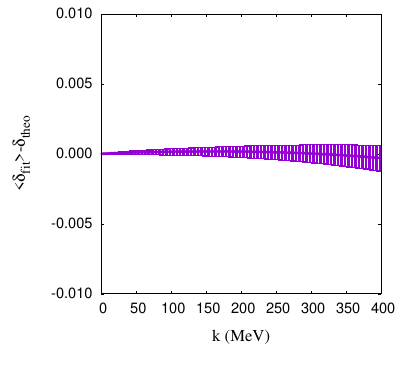}
\includegraphics[width=5 cm]{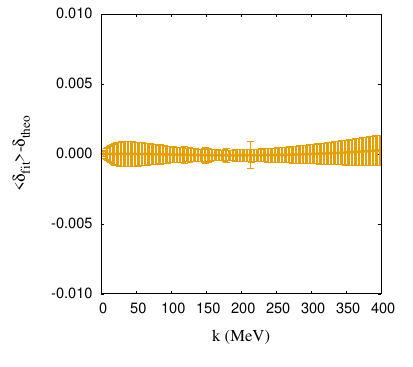}
\includegraphics[width=5 cm]{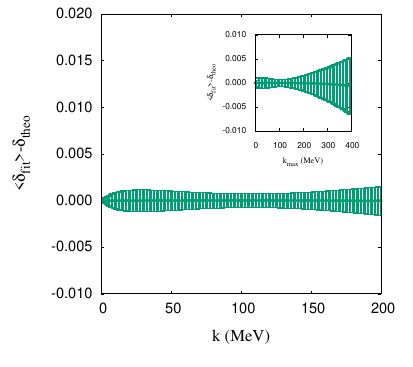}
\caption{\label{figdmean} Difference between the mean of
the 2000 fitted phase shifts and the exact result. The errors show the mean deviation of the fitted phase shifts. To compare, the scale is given by $\Delta \delta$, and the range by the range of the fitting. In the
right figure we also plot a larger range, showing the extrapolation up to 400 MeV. (Notice that the upper limit is doubled to include the extrapolation inset.) The color codes are those of Fig.~\ref{resV4}.}
\end{figure*}

\medskip
In Fig.~\ref{resV4} we can see that the residuals are normally distributed as expected. Here, we plot only three selected cases. Finally, in
Fig.~\ref{figdmean} we plot, for the same three cases, the difference between the mean of the 2000 fits and the exact result; the shown error stands for the mean deviation of the fits. 
As should be for a correct theory, this difference is always consistent with zero. In each case, the scale and range of the panels are  determined respectively by $\Delta \delta$ and $k_{\rm \max}$, 
so we can see that the mean deviation of the theory is smaller than the
precision of the data. In the right panel---N/D$_{22}$, $\Delta \delta=0.01^\circ$, $k_{\rm max}=200$ MeV---we can see that
the extrapolated phase shifts for $k>k_{\rm max}$ exhibit a mean deviation increasing with $k$, but still lower than $\Delta \delta$  in this energy region. Nevertheless, the extrapolation is still consistent with zero, as expected.

\subsection{LO theory}

\begin{figure*}[th]
\centering
\includegraphics[width=5 cm]{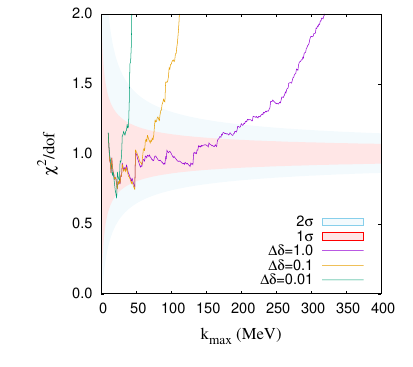}
\includegraphics[width=5 cm]{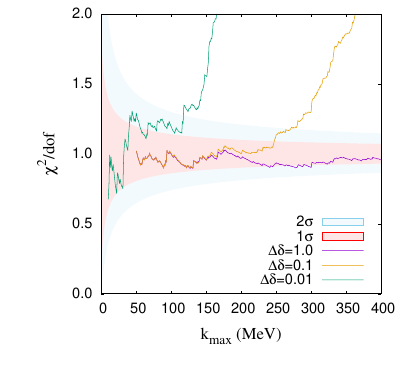}
\includegraphics[width=5 cm]{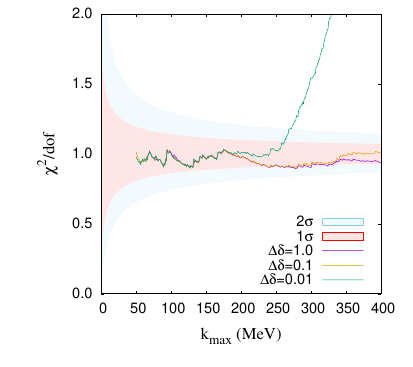}
\includegraphics[width=5 cm]{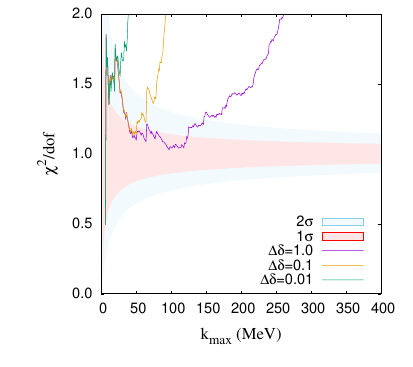}
\includegraphics[width=5 cm]{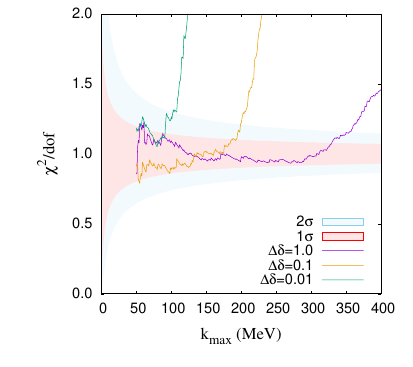}
\includegraphics[width=5 cm]{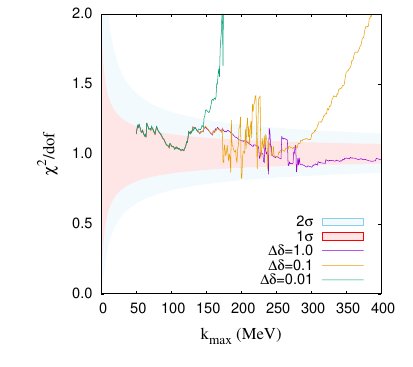}
\caption{\label{figV1} $\chi^2$/dof for fits with data between 1 MeV and $k_{\rm max}$. The red band shows the 1$\sigma$ confidence level, while
the blue band shows the 2$\sigma$ region. Data are generated as explained in the text with $\Delta\delta = 1^\circ$ (purple line), $0.1^\circ$ (gold line) and $0.01^\circ$ (green line). In all cases the theory at LO is considered. The upper row of panels corresponds to the N/D solutions (N/D$_{11}$, N/D$_{12}$, and N/D$_{22}$ from left to right). The lower row corresponds to the LS solutions (one, two, and three contact terms from left to right).}
\end{figure*}

We now consider calculations that include only $V_1(r)$---our `LO theory'. Again, the regularity of the potential allows us to make any number of subtractions, so following the previous section, we first compute $\chi^2/$dof for the N/D$_{11}$, N/D$_{12}$, and N/D$_{22}$ solutions with different precisions. The results are given in the panels of the first row in Fig.~\ref{figV1}. The second row corresponds to renormalizing with one, two, and three contact terms as explained in Section~\ref{Sct}.

\medskip
The first thing to note is that decreasing the precision of the experimental data (increasing the value of $\Delta \delta$) allows us to describe the data
at higher energies. Also, increasing the number of renormalization conditions allows us to reproduce the data at higher energies. This is, of course, what one would expect.

\medskip
The N/D$_{11}$ solution (one single renormalization condition) deviates very soon from the data. For example, if we take $\Delta\delta=1^\circ$, $\chi^2$/dof exceeds the
2$\sigma$ band already at $k_{\rm max}\approx200\,\text{MeV}$. 
For the case with contact terms, the results show the same pattern, but deviations start at smaller $k_{\rm max}$ as compared to the N/D case, also when $\Delta \delta$ is reduced, which is probably due to the small cutoff artifacts produced by the regulator function. 

\begin{figure*}[th]
\centering
\includegraphics[width=5 cm]{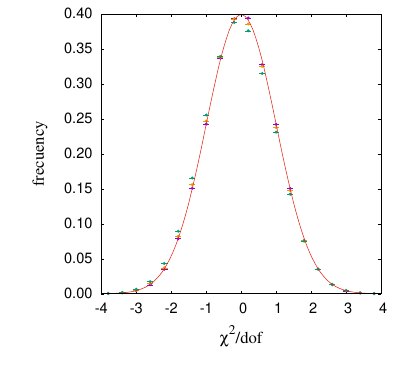}

\includegraphics[width=5 cm]{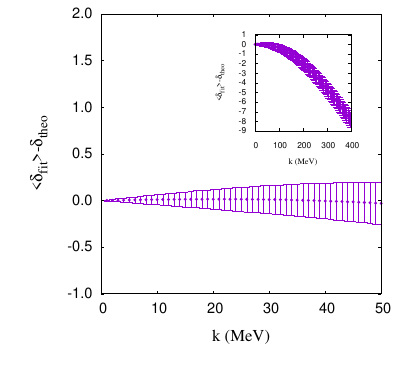}
\includegraphics[width=5 cm]{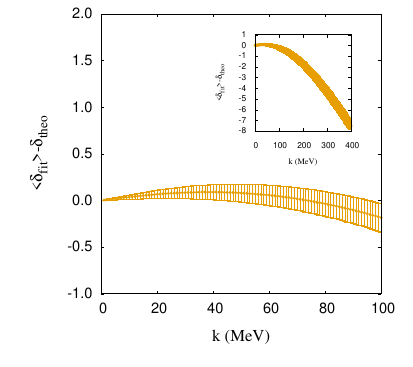}
\includegraphics[width=5 cm]{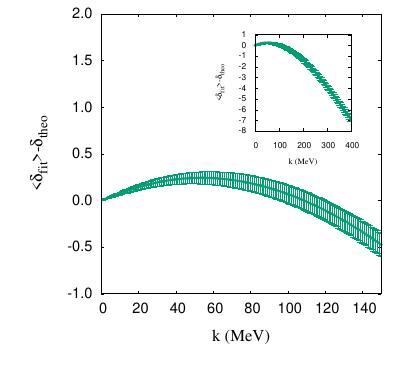}

\includegraphics[width=5 cm]{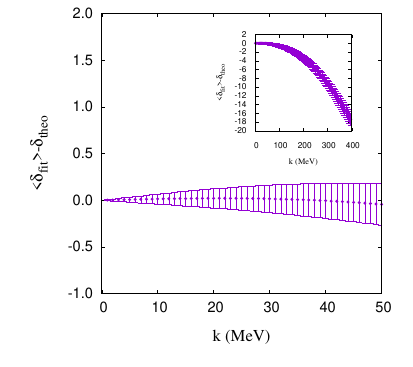}
\includegraphics[width=5 cm]{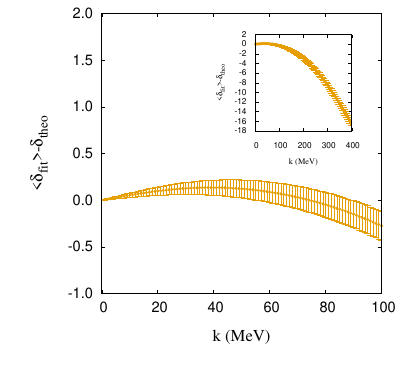}
\includegraphics[width=5 cm]{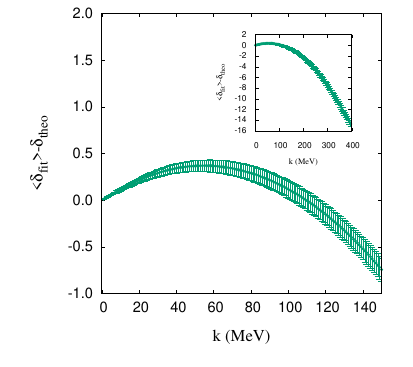}
\caption{\label{bootV1} The upper panel corresponds to the distribution of the residuals in three different cases for the N/D calculation. For the N/D$_{11}$ case with $\Delta\delta=1.0^\circ$, we give the results of $k_{\rm max}$ = 50 MeV (purple), 100 MeV (gold), and 150 MeV (green). The panels on the second row give the difference between the mean of the fits and the exact value of the theory with the same color code. The error is given by the mean deviation of the fits. Same for the third row of panels, but for the one-counterterm case.}
\end{figure*}

\medskip
The above is just the result for one particular experiment, so now we do a bootstrap of 2000 experiments for $k_{\rm max} = 50$, 100, and 150 MeV, having fixed $\Delta\delta=1.0^\circ$. Given the N/D$_{11}$ solution, Fig.~\ref{bootV1} shows the residuals (upper panel) and the deviations from the theory of the mean of the fits (second row of panels). The third row corresponds to renormalizing with one contact term.
The residuals are in good agreement with the standard normal for 50 MeV, and only slightly deviate for the other two cases.
However, we can see that the mean of the fits minus the exact value is clearly inconsistent with zero  for 
$k_{\rm max}=150$ MeV. 
Also, the extrapolation shows deviation of the order of $8^\circ$ at 400 MeV, as seen in the second row for the N/D calculation. 
The same is observed in the counterterm case, although deviations are stronger here.

\medskip
Table~\ref{V1S11} collects the corresponding $\chi^2/$dof. We verify a clear deviation for $k_{\rm max}=150$ MeV that is larger for the counterterm case. The values of the fitted scattering lengths as the mean of all the fits are given for the N/D case. We compare with the exact value given in Table \ref{EREV}, and see that the value obtained at 50 MeV is consistent, but the values at 100 and 150 MeV are 
{slightly and clearly} inconsistent, respectively, showing that the theory is not accurate enough.

\begin{table*}[th]
\caption{\label{V1S11} Bootstrap to 2000 experiments for the N/D$_{11}$ (upper table) and one-counterterm
(lower table) cases at LO and $\Delta \delta=1^\circ$.
The first column shows the maximum momentum considered in the fit, the second one gives $\chi^2/$dof of the bootstrap, the third one gives the result from the theoretical distribution, and the last one (for the N/D case) the value of the fitted scattering length.}
\centering
\begin{tabular}{ccccc}
\hline\noalign{\smallskip}
$k_{\rm max}$(MeV)   & $\chi^2/$dof  & $\chi^2/$dof (th.) & $a$ (fm) \\
\noalign{\smallskip}\hline\noalign{\smallskip}
 50 & $0.996 \pm 0.204$ & $0.986^{+0.214}_{-0.187}$ & $-0.619 \pm 0.019$ \\
\noalign{\smallskip}\hline\noalign{\smallskip}
100 & $1.009 \pm 0.145$ & $0.993^{+0.148}_{-0.135}$ & $-0.628 \pm 0.008$ \\
\noalign{\smallskip}\hline\noalign{\smallskip}
150 & $1.044 \pm 0.122$ & $0.996^{+0.120}_{-0.111}$ & $-0.642 \pm 0.005$ \\
\noalign{\smallskip}\hline
\end{tabular}

\medskip
\begin{tabular}{cccc}
\hline\noalign{\smallskip}
$k_{\rm max}$(MeV)   & $\chi^2/$dof  & $\chi^2/$dof (th.) \\ 
\noalign{\smallskip}\hline\noalign{\smallskip}
 50 & $0.993 \pm 0.193$ & $0.986^{+0.214}_{-0.187}$ \\ 
\noalign{\smallskip}\hline\noalign{\smallskip}
100 & $1.008 \pm 0.133$ & $0.993^{+0.148}_{-0.135}$ \\ 
\noalign{\smallskip}\hline\noalign{\smallskip}
150 & $1.090 \pm 0.124$ & $0.996^{+0.120}_{-0.111}$ \\ 
\noalign{\smallskip}\hline
\end{tabular}
\end{table*}

\begin{figure*}[th]
\centering
\includegraphics[width=5 cm]{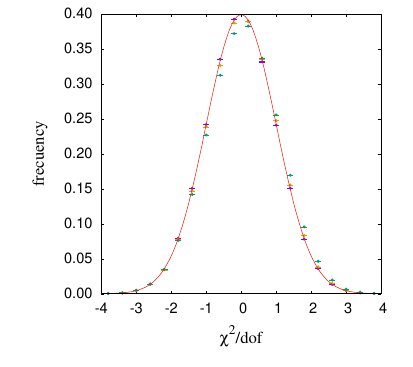}

\includegraphics[width=5 cm]{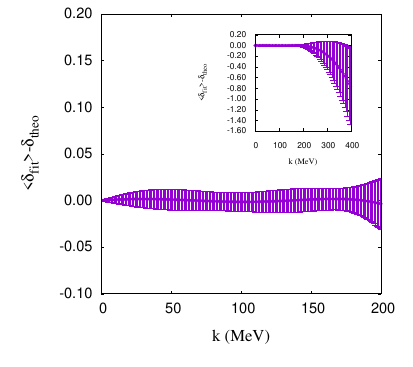}
\includegraphics[width=5 cm]{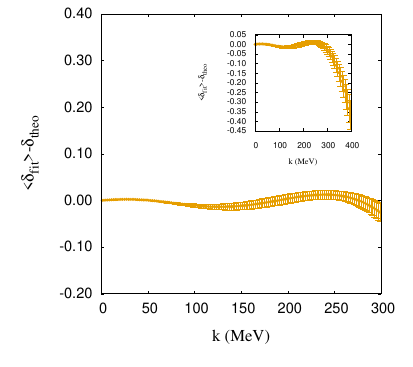}
\includegraphics[width=5 cm]{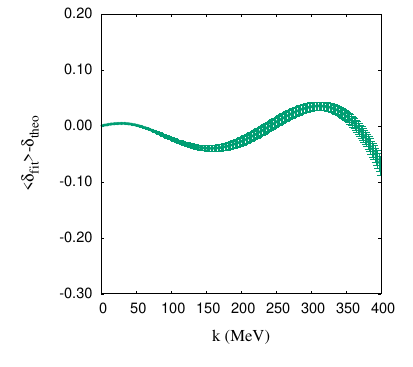}

\includegraphics[width=5 cm]{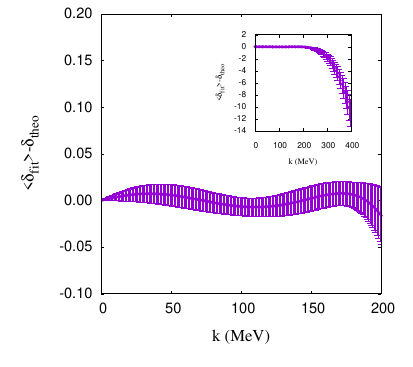}
\includegraphics[width=5 cm]{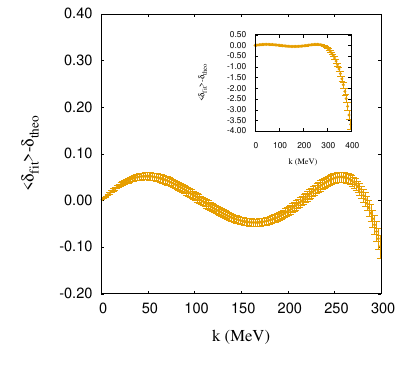}
\includegraphics[width=5 cm]{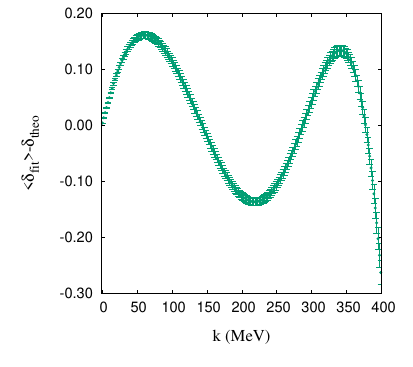}
\caption{\label{bootV1S22} {Same as in Fig. \ref{bootV1}, but for the N/D$_{22}$ or the three-counterterm solution, $\Delta\delta=0.1^\circ$, and $k_{\rm max}=200,\,300,\,400$ MeV.}}
\end{figure*}

\medskip
We now turn to the N/D$_{22}$ solution (three renormalization conditions). From the third panel of the upper row in Fig.~\ref{figV1}, we would expect to describe the data up to $k_{\rm max}=400$ MeV with $0.1^\circ$ and $1.0^\circ$ precision, but not with $0.01^\circ$. However, Fig.~\ref{figteo} gives us the exact deviation from the full theory, from where we learn that a $0.1^\circ$ precision allows us to reach $\approx 180$ MeV only. We test the $0.1^\circ$ case doing bootstraps to 2000 experiments for $k_{\rm max}=200$, $300$, and $400$ MeV. The results are shown in Fig.~\ref{bootV1S22}. The upper panel shows the distribution of residuals for the N/D case---a good agreement is observed for $k_{\rm max}=200$ MeV and slight deviations for the other two cases. In the lower panels, we show the difference between the fitted and the exact results, and we can see that for $k_{\rm max}=200$ MeV the result is consistent with zero. For 300 MeV there is still consistency, but for 400 MeV a clear statistical deviation is observed. Notice that the uncertainty of the fits is of the order or larger than $\Delta\delta$ in this case. Again, we see a stronger deviation for the renormalization with contact terms with respect to the N/D result.

\medskip
In Table~\ref{V1S22} we show $\chi^2/$dof and the fitted ERE parameters. We see that $\chi^2/$dof is already slightly deviated for 200 MeV, and the deviation increases in the other two cases. Recalling the values of the ERE parameters from the full theory that were given in Table \ref{EREV}, we confirm that the results obtained
from the 200 and 300 MeV cases are compatible with them, although the 300 MeV case shows a clear deviation. For the 400 MeV case, the
values obtained are not compatible at the 1$\sigma$ level. For the case with contact terms, we see that the deviations in $\chi^2/$dof exceed the ones for the N/D calculation.

\begin{table*}[th]
\caption{\label{V1S22} At LO, bootstrap to 2000 experiments for the N/D$_{22}$ case (upper table) and three contact terms (lower table), with $\Delta \delta=0.1^\circ$. The first column shows the maximum momentum considered in the fits, the second one gives $\chi^2/$dof of the bootstrap, the third one provides the result from the theoretical $\chi^2$ distribution, and the last columns display the values of the scattering length, the effective range and the shape parameter, respectively.}
\centering
\begin{tabular}{ccccccc}
\noalign{\smallskip}\hline\noalign{\smallskip}
$k_{\rm max}$(MeV)   & $\chi^2/$dof  & $\chi^2/$dof (th.) & $a$ (fm) & $r$ (fm) & $v_2$ (fm$^3$) \\
\noalign{\smallskip}\hline\noalign{\smallskip}
200 & $1.002 \pm 0.100$ & $0.997^{+0.104}_{-0.097}$ & $-0.6154 \pm 0.0013$ & $28.15 \pm 0.06$ & $19.05  \pm 0.10  $ \\
\noalign{\smallskip}\hline\noalign{\smallskip}
300 & $1.012 \pm 0.082$ & $0.998^{+0.084}_{-0.080}$ & $-0.61565 \pm 0.00008$ & $28.1568 \pm 0.0015$ & $19.16\pm 0.03$ \\
\noalign{\smallskip}\hline\noalign{\smallskip}
400 & $1.073 \pm 0.075$ & $0.998^{+0.073}_{-0.069}$ & $-0.61601 \pm 0.00004$ & $28.1583 \pm 0.0005$ & $19.264\pm 0.015$ \\
\noalign{\smallskip}\hline\noalign{\smallskip}
\end{tabular}

\begin{tabular}{cccc}
\noalign{\smallskip}\hline\noalign{\smallskip}
$k_{\rm max}$(MeV)   & $\chi^2/$dof  & $\chi^2/$dof (th.) \\ 
\noalign{\smallskip}\hline\noalign{\smallskip}
200 & $1.002 \pm 0.099$ & $0.997^{+0.104}_{-0.097}$ \\ 
\noalign{\smallskip}\hline\noalign{\smallskip}
300 & $1.122 \pm 0.088$ & $0.998^{+0.084}_{-0.080}$ \\ 
\noalign{\smallskip}\hline\noalign{\smallskip}
400 & $2.020 \pm 0.120$ & $0.998^{+0.073}_{-0.069}$ \\ 
\noalign{\smallskip}\hline\noalign{\smallskip}
\end{tabular}
\end{table*}

\begin{figure*}[th]
\centering
\includegraphics[width=8 cm]{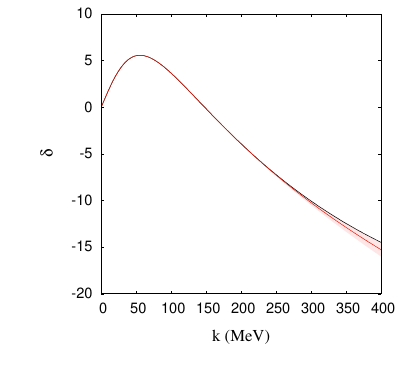}
\includegraphics[width=8 cm]{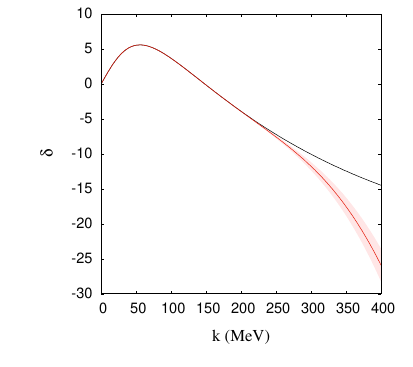}
\caption{\label{V1S22200} At LO, phase shift up to 400 MeV for the bootstrap fit from N/D$_{22}$ (left) and three contacts (right) extrapolated from $k_{\rm max}=200$ MeV and $\Delta\delta=0.1^\circ$. The red
line shows the mean of the fits and the red shaded area is the 1$\sigma$ confidence level. For comparison, the black line provides the result from the full theory.}
\end{figure*}

\medskip
In Fig.~\ref{bootV1S22} we see that the fits are in agreement with the full theory in the fitting region for the 200 MeV case, but
the extrapolation shows a large uncertainty area. We show in Fig.~\ref{V1S22200} the extrapolation up to 400 MeV of the fit
with $k_{\rm max}=200$ MeV. For the N/D case we are still close to the exact result, while for the contact terms we see a clear deviation.
Again, this seems to be due to cutoff artifacts.

\subsection{NLO theory}

\begin{figure*}
\centering
\includegraphics[width=6 cm]{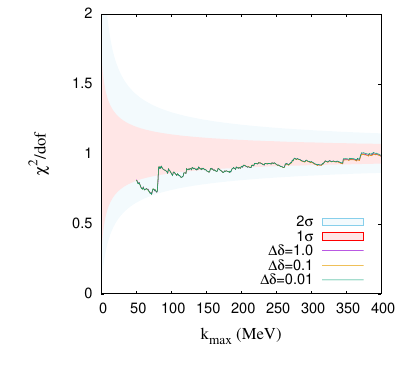}
\includegraphics[width=6 cm]{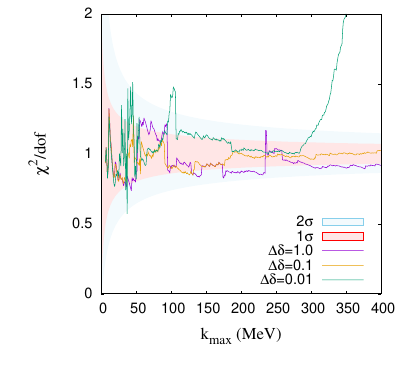}
\caption{\label{fig2} $\chi^2/$dof for fits with data between 1 MeV and $k_{\rm max}$. The red band shows the 1$\sigma$ confidence level, while
the blue band shows the 2$\sigma$ region. Data are generated as explained in the text with $\Delta\delta = 1^\circ$ (purple line), $0.1^\circ$ (gold line) and $0.01^\circ$ (green line). The left figure corresponds to the N/D$_{22}$ solution
at NLO, while the right figure with three contact terms at NLO.}
\end{figure*}

At NLO we include the singular repulsive potential $V_2$, which is compatible with none and three renormalization conditions, as in the N/D$_{01}$ and N/D$_{22}$ solutions, respectively. 
So we consider the N/D$_{22}$ solution and renormalization with three contact terms.
In Fig.~\ref{fig2} we show $\chi^2/$dof for one experiment
with three different values of $\Delta\delta$, on the left for the N/D result and on the right
for three contact terms. We can see that the results are inside the 2$\sigma$ confidence level up to 400 MeV for the N/D case. With three contact terms, only above 300 MeV does the result go out of the
2$\sigma$ confidence level for the most precise calculation with $\Delta \delta=0.01^\circ$. 
This indicates that the theory is accurate up to this energy. From Fig.~\ref{figteo}, we should expect a precise
theory up to 400 MeV for $0.1^\circ$ and up to 300 MeV for $0.01^\circ$ for the N/D calculation.

\begin{figure*}
\centering
\includegraphics[width=5 cm]{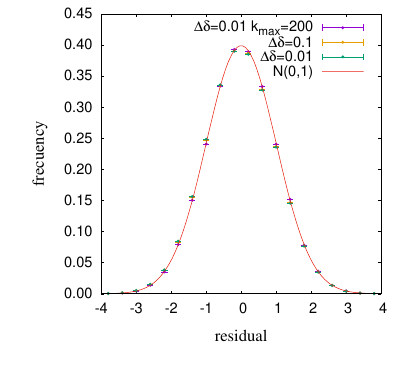}

\includegraphics[width=5 cm]{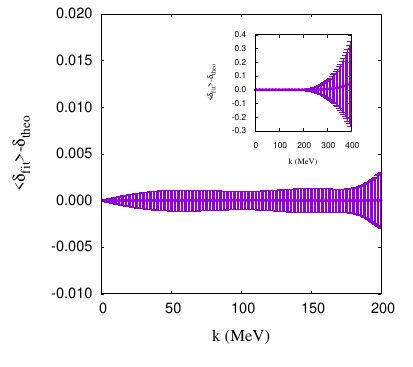}
\includegraphics[width=5 cm]{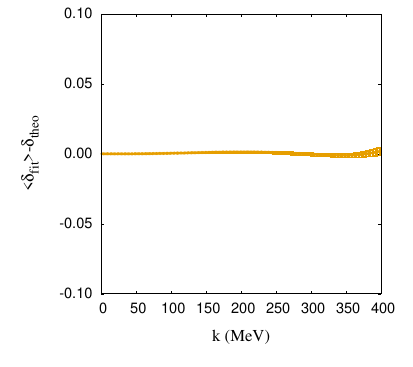}
\includegraphics[width=5 cm]{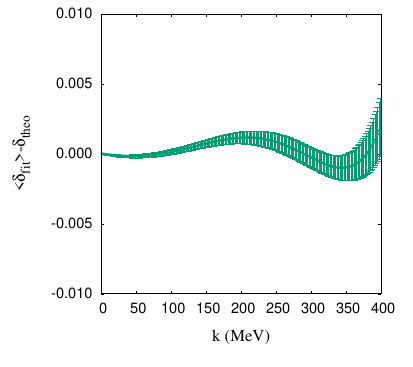}

\includegraphics[width=5 cm]{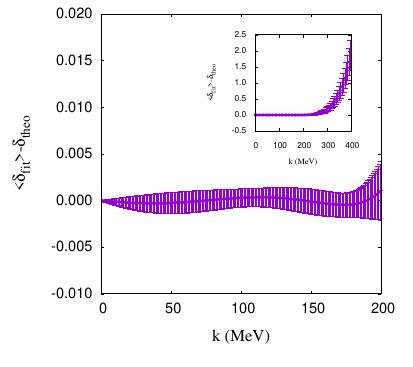}
\includegraphics[width=5 cm]{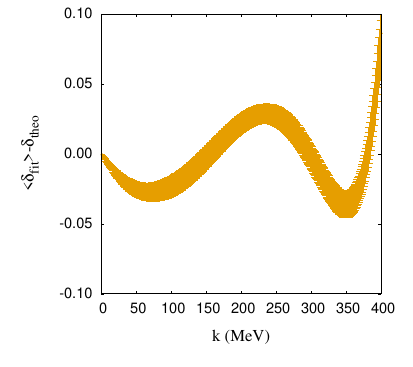}
\includegraphics[width=5 cm]{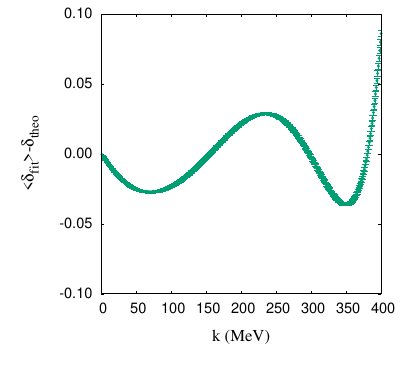}
\caption{\label{bootV2} {Same as in Fig. \ref{bootV1}, now after the inclusion of $V_2$, for the N/D$_{22}$ or the three-counterterm solution, with $\Delta\delta=0.01^\circ$, $k_{\rm max}=200$ MeV; $\Delta\delta=0.1^\circ$, $k_{\rm max}=400$ MeV; $\Delta\delta=0.01^\circ$, $k_{\rm max}=400$ MeV.}} 
\end{figure*}

\medskip
As in previous cases, we now do a bootstrap with 2000 experiments, and show the results in Fig.~\ref{bootV2}. We consider fits for
$k_{\rm max}=200$ MeV with $\Delta \delta=0.01^\circ$, and for $400$ MeV with $0.1^\circ$ and $0.01^\circ$. 
In the upper panel we show the residuals for the N/D calculation. The distribution is almost perfect for the fit with $k_{\rm max}=200$ MeV,
while for the other cases we find small deviations.
In the panels on the second row
we can see that the uncertainty of the theory is smaller than the one of the data for the N/D case. 
For the 200 MeV fit, the difference shown in the
figure is perfectly consistent with zero, which ensures that we have a precise description of the data. In the 400 MeV case, we can see a small deviation consistent with Fig.~\ref{figteo}.
Finally, the third row of panels shows results with three contact terms. The fit with $k_{\rm max}=200$ MeV is very
similar to the previous case, 
although the extrapolation is worse. However, for the fits with $k_{\rm max}=400$ MeV we see that the difference between the mean of the fits and the theoretical result is not compatible with zero. The last panel shows that this discrepancy is one order of magnitude larger than the precision of the data, implying that, in this case, the theory is not accurate enough for these data.

\medskip
In Table~\ref{TV2S22} we give $\chi^2/$dof and ERE parameters in the three cases considered for the N/D method (upper table) and with three contact terms (lower table). Regarding N/D, we can see a slight deviation in $\chi^2/$dof for the 400 MeV fits, although the standard deviation is in agreement with the theoretical one. However, we confirm a much more precise determination of the ERE parameters. For the 200 MeV case we observe perfect agreement with the values of the full theory. Small deviations are seen in both 400 MeV fits. 

\medskip
For three contact terms, we see stronger deviations, where $k_{\rm max}=400$ MeV and $\Delta \delta=0.01^\circ$ is the worst case, as expected. This shows that, although the theory is almost accurate enough for data with an error $\Delta \delta=0.1^\circ$, it is clearly not accurate enough for $\Delta \delta=0.01^\circ$.

\begin{table*}
\caption{\label{TV2S22} At NLO, bootstrap to 2000 experiments for the N/D$_{22}$ case (upper table) and three contact terms (lower table). The first column shows the maximum momentum of data considered, the second one gives the value of $\Delta\delta$, the third one provides $\chi^2/$dof of the bootstrap, the fourth one gives
the result from the theoretical distribution, and the last columns the values of the ERE parameters from the bootstrap in the N/D case.}
\centering
\begin{tabular}{cccccccc}
\noalign{\smallskip}\hline\noalign{\smallskip}
$k_{\rm max}$(MeV)   & $\Delta \delta$ & $\chi^2/$dof  & $\chi^2/$dof (th.) & $a$ (fm) & $r$ (fm) & $v_2$ (fm$^3$) \\
\noalign{\smallskip}\hline\noalign{\smallskip}
200 & $0.01^\circ$ & $1.003 \pm 0.101$ & $0.997^{+0.104}_{-0.097}$ & $-0.61518 \pm 0.00012$ & $28.148 \pm 0.006$ & $19.007 \pm 0.009$ \\
\noalign{\smallskip}\hline\noalign{\smallskip}
400 & $0.1^\circ$ & $1.010 \pm 0.070$ & $0.998^{+0.073}_{-0.069}$ & $-0.615174 \pm 0.000005$ & $28.1481 \pm 0.0003$ & $19.0036 \pm 0.0012$ \\
\noalign{\smallskip}\hline\noalign{\smallskip}
400 & $0.01^\circ$ & $1.011 \pm 0.070$ & $0.998^{+0.073}_{-0.069}$ & $-0.615158 \pm 0.000010$ & $28.1492 \pm 0.0007$ & $19.0059 \pm 0.0015$ \\
\noalign{\smallskip}\hline\noalign{\smallskip}
\end{tabular}

\begin{tabular}{ccccc}
\noalign{\smallskip}\hline\noalign{\smallskip}
$k_{\rm max}$(MeV)   & $\Delta \delta$ & $\chi^2/$dof  & $\chi^2/$dof (th.) \\
\noalign{\smallskip}\hline\noalign{\smallskip}
200 & $0.01^\circ$ & $1.001 \pm 0.100$ & $0.997^{+0.104}_{-0.097}$ \\
\noalign{\smallskip}\hline\noalign{\smallskip}
400 & $0.1^\circ$ & $1.053 \pm 0.075$ & $0.998^{+0.073}_{-0.069}$ \\
\noalign{\smallskip}\hline\noalign{\smallskip}
400 & $0.01^\circ$ & $6.434 \pm 0.247$ & $0.998^{+0.073}_{-0.069}$ \\
\noalign{\smallskip}\hline\noalign{\smallskip}
\end{tabular}
\end{table*}

\begin{figure*}
\centering
\includegraphics[width=6 cm]{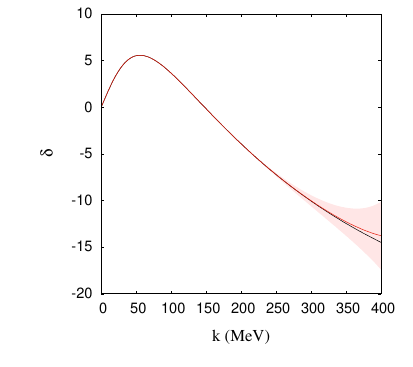}
\includegraphics[width=6 cm]{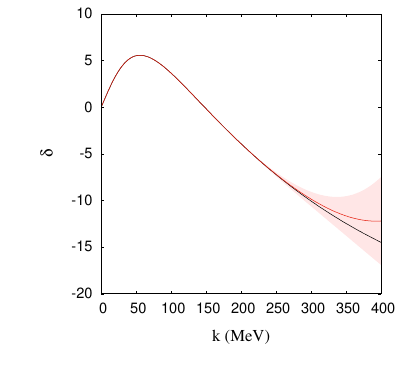}

\includegraphics[width=6 cm]{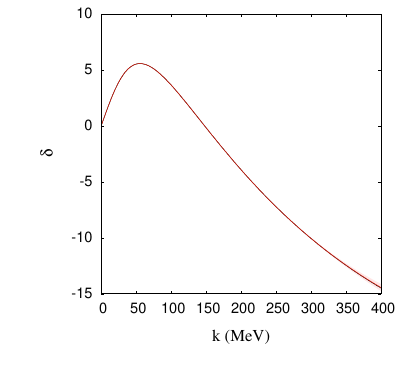}
\includegraphics[width=6 cm]{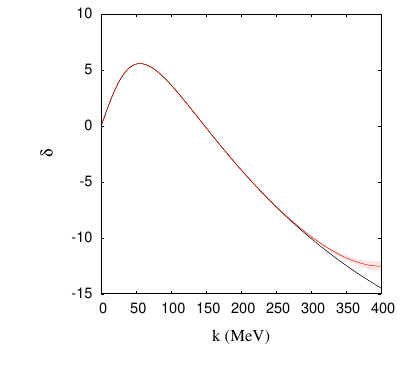}
\caption{\label{V2S22200} At NLO, phase shift up to 400 MeV extrapolated from the $k_{\rm max}=200$ MeV fit.
The left upper (left lower) panel corresponds to the N/D$_{22}$ solution with $\Delta\delta=0.1^\circ$ ($\Delta\delta=0.01^\circ$); analogously for the right panels obtained from the three-counterterm case. The red line shows the mean of the fits and the red shaded area is the 1$\sigma$ confidence level. The black line stands for the full-theory result.}
\end{figure*}

\medskip
Finally, in Fig.~\ref{V2S22200} we show the phase shift up to 400 MeV extrapolated from the fit with $k_{\rm max}=200$ MeV. If we compare with Fig.~\ref{V1S22200}, we can see that NLO makes a much better extrapolation and a much better precision of the theory with respect to the LO case with the same renormalization conditions. We can also see that the extrapolation gets worse in the case of three contact terms.

\section{\texorpdfstring{{Application to the $\boldsymbol{^1S_0}$ $\boldsymbol{NN}$ partial wave in chiral EFT}}{Application to the 1S0 NN partial wave in chiral EFT}}
\label{chiEFT}

In this section, we apply the N/D method to study the $^1S_0$ $NN$ partial wave in chiral EFT. At LO,
the potential is given by the charge-dependent OPE. At NLO, we have to include TPE
with vertices from the axial-vector coupling and the Weinberg-Tomozawa term. Notice that TPE at NLO does not encode any new parameter. 

\medskip
As mentioned previously, here we do not invoke any power counting on the number of subtractions, but only a hierarchy of long-range interactions. In order to have an accurate description we do three subtractions both at LO and NLO (N/D$_{22}$ solution), thus fixing $a$, $r$, and $v_2$.

\medskip
For the data we use the $np$ Granada phase-shift analysis {\cite{Rodrigoprivate,NavarroPerez:2013usk,PhysRevC.88.064002}}, which is an energy-dependent analysis. For the bootstrap, we generate random values with a Gaussian distribution centered on the Granada value and a standard deviation given by the error. We use data from 1 MeV to $k_{\rm max}$ in 1 MeV bins as in the toy model.

\begin{table*}
\caption{\label{cchiPTV2S22} Bootstrap to 2000 experiments for the N/D$_{22}$ solution at LO and NLO. The first column shows the maximum momentum of data considered, the second one indicates the order, the third and fourth one give $\chi^2/$dof from the bootstrap and from the theoretical distribution, and the last three columns provide the values of the ERE parameters from the bootstrap.}
\centering
\begin{tabular}{cccccccc}
\noalign{\smallskip}\hline\noalign{\smallskip}
$k_{\rm max}$(MeV)   & order & $\chi^2/$dof  & $\chi^2/$dof (th.) & $a$ (fm) & $r$ (fm) & $v_2$ (fm$^3$) \\
\noalign{\smallskip}\hline\noalign{\smallskip}
100 & LO & $1.003 \pm 0.146$ & $0.993^{+0.150}_{-0.136}$ & $-23.7352 \pm 0.0020$ & $2.6741 \pm 0.0015$ & $-0.4952 \pm 0.0049$ \\
\noalign{\smallskip}\hline\noalign{\smallskip}
150 & LO & $1.004 \pm 0.118$ & $0.995^{+0.121}_{-0.112}$ & $-23.7351 \pm 0.0019$ & $2.6739 \pm 0.0011$ & $-0.4947 \pm 0.0015$ \\
\noalign{\smallskip}\hline\noalign{\smallskip}
200 & LO & $1.057 \pm 0.106$ & $0.997^{+0.104}_{-0.097}$ & $-23.7341 \pm 0.0019$ & $2.6721 \pm 0.0010$ & $-0.4915 \pm 0.0008$ \\
\noalign{\smallskip}\hline\noalign{\smallskip}
250 & LO & $1.48  \pm 0.12 $ & $0.997^{+0.093}_{-0.087}$ & $-23.7315 \pm 0.0019$ & $2.6678 \pm 0.0009$ & $-0.4870 \pm 0.0006$ \\
\noalign{\smallskip}\hline\noalign{\smallskip}
300 & LO & $3.02  \pm 0.18 $ & $0.998^{+0.084}_{-0.080}$ & $-23.7269 \pm 0.0019$ & $2.6601 \pm 0.0008$ & $-0.4831 \pm 0.0006$ \\
\noalign{\smallskip}\hline\noalign{\smallskip}
100 & NLO& $1.005 \pm 0.146$ & $0.993^{+0.150}_{-0.136}$ & $-23.7354 \pm 0.0020$ & $2.6744 \pm 0.0015$ & $-0.5003 \pm 0.0053$ \\
\noalign{\smallskip}\hline\noalign{\smallskip}
150 & NLO& $1.011 \pm 0.119$ & $0.995^{+0.121}_{-0.112}$ & $-23.7359 \pm 0.0019$ & $2.6754 \pm 0.0012$ & $-0.5047 \pm 0.0018$ \\
\noalign{\smallskip}\hline\noalign{\smallskip}
200 & NLO& $1.011 \pm 0.101$ & $0.997^{+0.104}_{-0.097}$ & $-23.7361 \pm 0.0019$ & $2.6758 \pm 0.0010$ & $-0.5060 \pm 0.0010$ \\
\noalign{\smallskip}\hline\noalign{\smallskip}
250 & NLO & $1.034 \pm 0.093$ & $0.997^{+0.093}_{-0.087}$ & $-23.7356 \pm 0.0019$ & $2.6750 \pm 0.0009$ & $-0.5047 \pm 0.0007$ \\
\noalign{\smallskip}\hline\noalign{\smallskip}
300 & NLO & $1.228 \pm 0.098$ & $0.998^{+0.084}_{-0.080}$ & $-23.7340 \pm 0.0019$ & $2.6723 \pm 0.0009$ & $-0.5020 \pm 0.0006$ \\
\noalign{\smallskip}\hline\noalign{\smallskip}
\end{tabular}
\end{table*}

\begin{figure*}
\centering
\includegraphics[width=6 cm]{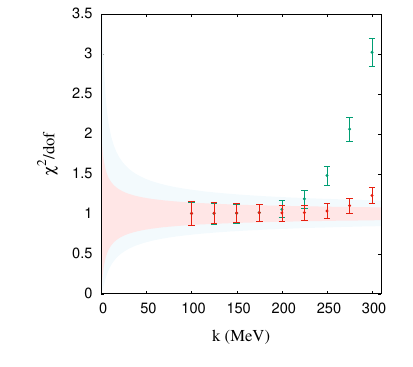}
\caption{\label{chi2CD} $\chi^2/$dof for the bootstraps to the Granada phase shift analysis at LO
(green points) and NLO (red points).}
\end{figure*}

\medskip
In Table~\ref{cchiPTV2S22} we give $\chi^2/$dof and ERE parameters for different $k_{\rm max}$; $\chi^2/$dof is also shown in Fig.~\ref{chi2CD}. The figure shows that the LO case is statistically consistent with the data up to $k_{\rm max}$ of the order of 175 MeV, while the NLO case increases it up to 225--250 MeV.

\begin{figure*}
\centering
\includegraphics[width=6 cm]{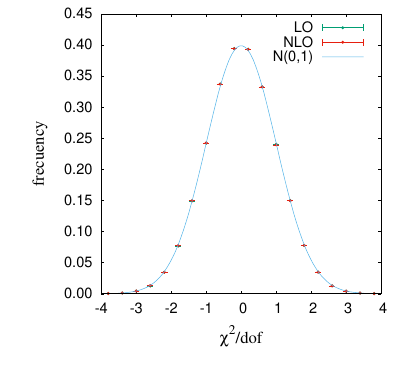}
\includegraphics[width=6 cm]{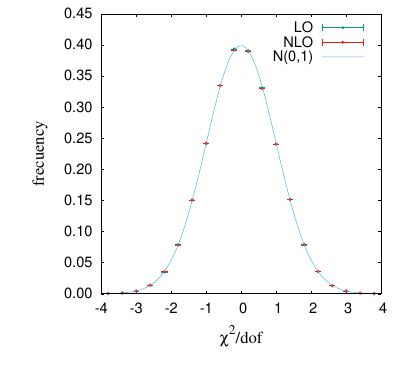}
\caption{\label{chiPTCDresid} Residuals for the bootstrap fits corresponding to the N/D$_{22}$ case for $k_{\rm max}=100$ MeV (left) and $k_{\rm max}=150$ MeV (right). The green (red) points refer to LO (NLO), and the blue line shows the expected distribution.}
\end{figure*} 

\begin{figure*}
\centering
\includegraphics[width=6 cm]{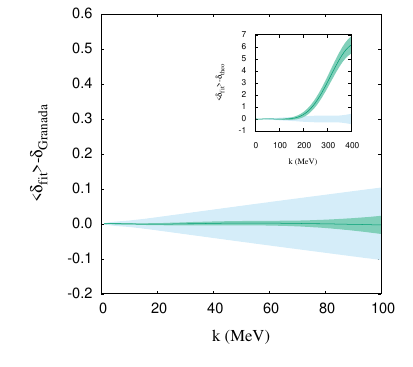}
\includegraphics[width=6 cm]{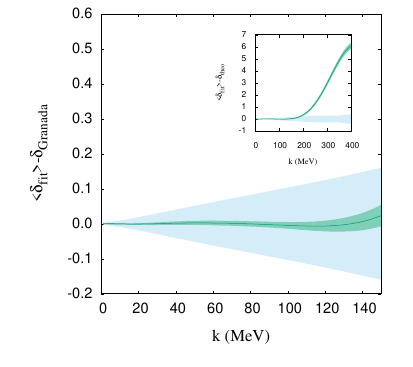}
\includegraphics[width=6 cm]{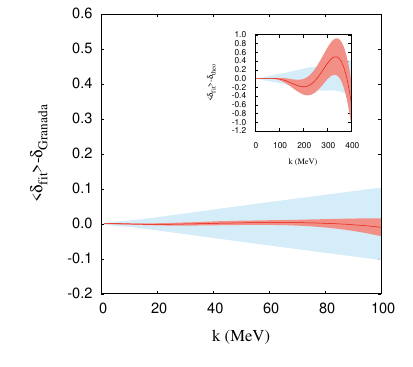}
\includegraphics[width=6 cm]{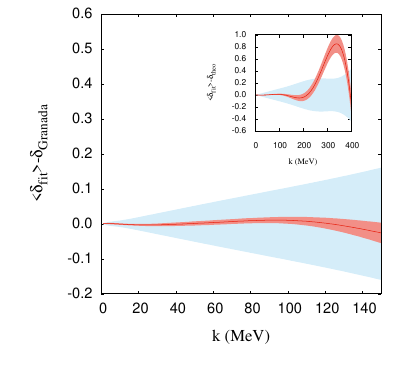}
\caption{\label{chiPTCDmean} Bootstrap fits corresponding to the N/D$_{22}$ solution for $k_{\rm max}=100$ MeV (left column) and $k_{\rm max}=150$ MeV (right column).
The first row shows in a green line the difference between the mean of the fits at LO and the Granada phase shifts, the blue shaded area is the 1$\sigma$ confidence level of the Granada data, and the green shaded area is the 1$\sigma$ confidence level of the fits. Same in the second row in red for the NLO case.
}
\end{figure*} 

\medskip
In Fig.~\ref{chiPTCDresid} we show the residuals for the fits with
$k_{\rm max}=100$ MeV (left) and $k_{\rm max}=150$ MeV (right). 
In Fig.~\ref{chiPTCDmean} we show the difference between the mean of the fits and the Granada data for the same values of $k_{\rm max}$. As shown in Table~\ref{cchiPTV2S22}, $\chi^2/$dof is in perfect agreement with the theoretical distribution in both cases. As expected for both cases, the uncertainty of the fits is smaller than the data error and almost compatible with zero in the whole range.

\begin{figure*}
\centering
\includegraphics[width=6 cm]{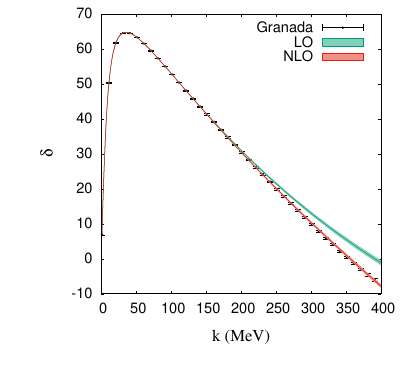}
\includegraphics[width=6 cm]{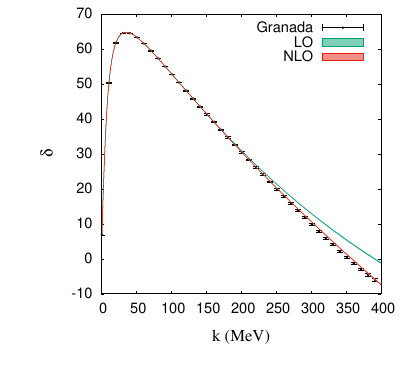}
\caption{\label{chiPTCDextp} 
Phase-shift extrapolation up to 400 MeV for the bootstrap fits associated to the N/D$_{22}$ solution for $k_{\rm max}=100$ MeV (left) and $k_{\rm max}=150$ MeV (right).
The solid dots with error bars correspond to the fitted data  from the Granada database (shown on 10 MeV intervals instead of 1 MeV for clarity). The shaded areas are the 1$\sigma$ confidence area of the fits. The green (red) area corresponds to the LO (NLO) case. 
}

\end{figure*} 

\medskip
Finally, in Fig.~\ref{chiPTCDextp} we show the extrapolation up to 400 MeV for fits with $k_{\rm max}=100$ and 150 MeV. The uncertainty of the extrapolated fits decreases for higher $k_{\rm max}$. In addition, the extrapolation agrees better with the Granada data for NLO than for LO.

\section{Conclusions}
\label{conc}

We have considered theories with an expansion in finite-range interactions supplemented by short-range contributions. The unknown short-range part is considered in two different frameworks. The first one takes a regulator function with a finite cutoff and contact terms that mimic the short-range interaction, allowing us to include singular pieces in the potential expansion; this is the usual way in which chiral EFT potentials are constructed. The second approach we use is the exact N/D method with subtractions, where the subtractions take into account the unresolved short-range physics. The latter method also allows us to include singular interactions; however, in this case no regularization is needed. For what concerns $S$ waves, the case of singular interactions without or with one subtraction is equivalent to the former framework without or with one contact term sending the cutoff to infinity. The N/D method allows us to include more subtractions, which is not possible in the infinite-cutoff limit. In both frameworks, the unknown short-range information is obtained by fitting experimental information.

\medskip
We have used the bootstrap technique to obtain distributions of the unknown short-range parameters and test the statistical consistency of the fits. As expected, decreasing the precision (increasing $\Delta \delta$) allows one to give a consistent description at higher energies. Including more renormalization conditions with more short-range parameters also allows one to go to higher energies. In addition, including more physics through the inclusion of next terms in the long-range expansion of the potential allows one to go to higher energies. In general, N/D and contact terms give similar results although the N/D method has been shown to be more accurate, probably due to the absence of cutoff artifacts present in the finite-cutoff calculations.

\medskip
We have also applied the bootstrap strategy to the study of the $^1 S_0$ $NN$ partial wave. The data are generated with the Granada phase-shift analysis, and the theories considered are the LO and NLO potentials in chiral EFT with three renormalization conditions (N/D$_{22}$). With the precision given by the Granada analysis, we have found that the N/D method is consistent with LO up to $k_{\rm max}\approx 175$ MeV, and with NLO up to $k_{\rm max}\approx 225-250$ MeV. Considering fittings up to $k_{\rm max} = $ 100 and 150 MeV where $\chi^2/$dof is in perfect agreement with the theoretical distribution, we see that the LO extrapolation deviates from the Granada phase shifts above 200 MeV, while the NLO extrapolation displays much better agreement.

\bigskip
\bigskip
\section*{Acknowledgements}
D.R.E. thanks R. Navarro Pérez for sharing with him the Granada phase shifts in 1 MeV bins. This work has been partially funded by Grant no.\ PID2022-141910NB-I00 funded by MCIN/AEI/\allowbreak 10.13039/\allowbreak 501100011033/ and Junta de Castilla y León under Grant no. SA091P24.

\bigskip
\bigskip
\bibliographystyle{apsrev4-1}       

\bibliography{bootstrap}   

@article{Machleidt:2011zz,
    author = "Machleidt, R. and Entem, D. R.",
    title = "{Chiral effective field theory and nuclear forces}",
    eprint = "1105.2919",
    archivePrefix = "arXiv",
    primaryClass = "nucl-th",
    journal = "Phys. Rept.",
    volume = "503",
    pages = "1--75",
    year = "2011"
}

@article{Epelbaum:2008ga,
    author = "Epelbaum, Evgeny and Hammer, Hans-Werner and Meissner, Ulf-G.",
    title = "{Modern Theory of Nuclear Forces}",
    eprint = "0811.1338",
    archivePrefix = "arXiv",
    primaryClass = "nucl-th",
    reportNumber = "HISKP-TH-08-18, FZJ-IKP-TH-2008-20",
    journal = "Rev. Mod. Phys.",
    volume = "81",
    pages = "1773--1825",
    year = "2009"
}

@article{Gasparyan:2021edy,
    author = "Gasparyan, A. M. and Epelbaum, E.",
    title = "{Nucleon-nucleon interaction in chiral effective field theory with a finite cutoff: Explicit perturbative renormalization at next-to-leading order}",
    eprint = "2110.15302",
    archivePrefix = "arXiv",
    primaryClass = "nucl-th",
    journal = "Phys. Rev. C",
    volume = "105",
    number = "2",
    pages = "024001",
    year = "2022"
}

@article{Gasparyan:2022isg,
    author = "Gasparyan, A. M. and Epelbaum, E.",
    title = "{{\textquotedblleft}Renormalization-group-invariant effective field theory{\textquotedblright} for few-nucleon systems is cutoff dependent}",
    eprint = "2210.16225",
    archivePrefix = "arXiv",
    primaryClass = "nucl-th",
    journal = "Phys. Rev. C",
    volume = "107",
    number = "3",
    pages = "034001",
    year = "2023"
}

@article{Baru:2019ndr,
    author = "Baru, V. and Epelbaum, E. and Gegelia, J. and Ren, X. -L.",
    title = "{Towards baryon-baryon scattering in manifestly Lorentz-invariant formulation of SU(3) baryon chiral perturbation theory}",
    eprint = "1905.02116",
    archivePrefix = "arXiv",
    primaryClass = "nucl-th",
    journal = "Phys. Lett. B",
    volume = "798",
    pages = "134987",
    year = "2019"
}

@article{NavarroPerez:2014bca,
    author = "Navarro P{\'e}rez, R. and Amaro, J. E. and Ruiz Arriola, E.",
    title = "{Bootstrapping the statistical uncertainties of NN scattering data}",
    eprint = "1407.3937",
    archivePrefix = "arXiv",
    primaryClass = "nucl-th",
    journal = "Phys. Lett. B",
    volume = "738",
    pages = "155--159",
    year = "2014"
}

@misc{Rodrigoprivate,
  author       = {Navarro P{\'e}rez, R.},
  title        = {Private communication}
}

@article{Yang:2021vxa,
    author = {Yang, C. -J. and Ekstr{\"o}m, A. and Forss{\'e}n, C. and Hagen, G. and Rupak, G. and van Kolck, U.},
    title = "{The importance of few-nucleon forces in chiral effective field theory}",
    eprint = "2109.13303",
    archivePrefix = "arXiv",
    primaryClass = "nucl-th",
    journal = "Eur. Phys. J. A",
    volume = "59",
    number = "10",
    pages = "233",
    year = "2023"
}

@article{Peng:2024aiz,
  title = {Contact operators in renormalization of attractive singular potentials},
  author = {Peng, Rui and Long, Bingwei and Xu, Fu-Rong},
  journal = {Phys. Rev. C},
  volume = {110},
  issue = {5},
  pages = {054001},
  numpages = {8},
  year = {2024},
  month = {Nov},
  publisher = {American Physical Society},
}

@article{vanKolck:2020llt,
    author = "van Kolck, U.",
    title = "{The Problem of Renormalization of Chiral Nuclear Forces}",
    eprint = "2003.06721",
    archivePrefix = "arXiv",
    primaryClass = "nucl-th",
    journal = "Front. in Phys.",
    volume = "8",
    pages = "79",
    year = "2020"
}

@article{Epelbaum:2017byx,
    author = "Epelbaum, Evgeny and Gegelia, Jambul and Mei\ss{}ner, Ulf-G",
    title = "{Wilsonian renormalization group versus subtractive renormalization in effective field theories for nucleon\textendash{}nucleon scattering}",
    eprint = "1705.02524",
    archivePrefix = "arXiv",
    primaryClass = "nucl-th",
    journal = "Nucl. Phys. B",
    volume = "925",
    pages = "161--185",
    year = "2017"
}

@article{Epelbaum:2017tzp,
    author = "Epelbaum, E. and Gegelia, J. and Mei\ss{}ner, Ulf-G.",
    title = "{Wilsonian renormalization group and the Lippmann-Schwinger equation with a multitude of cutoff parameters}",
    eprint = "1710.04178",
    archivePrefix = "arXiv",
    primaryClass = "nucl-th",
    journal = "Commun. Theor. Phys.",
    volume = "69",
    number = "3",
    pages = "303",
    year = "2018"
}

@article{Epelbaum:2018zli,
    author = "Epelbaum, E. and Gasparyan, A. M. and Gegelia, J. and Mei\ss{}ner, Ulf-G.",
    title = "{How (not) to renormalize integral equations with singular potentials in effective field theory}",
    eprint = "1810.02646",
    archivePrefix = "arXiv",
    primaryClass = "nucl-th",
    journal = "Eur. Phys. J. A",
    volume = "54",
    number = "11",
    pages = "186",
    year = "2018"
}

@article{Epelbaum:2019msl,
    author = "Epelbaum, E. and Gasparyan, A. M. and Gegelia, J. and Mei\ss{}ner, Ulf-G",
    title = "{Reply to ''Comment on ''How (not) to renormalize integral equations with singular potentials in effective field theory''}",
    eprint = "1903.01273",
    archivePrefix = "arXiv",
    primaryClass = "nucl-th",
    journal = "Eur. Phys. J. A",
    volume = "55",
    pages = "56",
    year = "2019"
}

@article{Furnstahl:2015rha,
    author = "Furnstahl, R. J. and Klco, N. and Phillips, D. R. and Wesolowski, S.",
    title = "{Quantifying truncation errors in effective field theory}",
    eprint = "1506.01343",
    archivePrefix = "arXiv",
    primaryClass = "nucl-th",
    journal = "Phys. Rev. C",
    volume = "92",
    number = "2",
    pages = "024005",
    year = "2015"
}

@article{Melendez:2017phj,
    author = "Meléndez, J. A. and Wesolowski, S. and Furnstahl, R. J.",
    title = "{Bayesian truncation errors in chiral effective field theory: nucleon-nucleon observables}",
    eprint = "1704.03308",
    archivePrefix = "arXiv",
    primaryClass = "nucl-th",
    journal = "Phys. Rev. C",
    volume = "96",
    number = "2",
    pages = "024003",
    year = "2017"
}

@article{Melendez:2019izc,
    author = "Meléndez, J. A. and Furnstahl, R. J. and Phillips, D. R. and Pratola, M. T. and Wesolowski, S.",
    title = "{Quantifying Correlated Truncation Errors in Effective Field Theory}",
    eprint = "1904.10581",
    archivePrefix = "arXiv",
    primaryClass = "nucl-th",
    journal = "Phys. Rev. C",
    volume = "100",
    number = "4",
    pages = "044001",
    year = "2019"
}

@article{Oller:2014uxa,
    author = "Oller, J. A.",
    title = "{Nucleon-Nucleon scattering from dispersion relations: next-to-next-to-leading order study}",
    eprint = "1402.2449",
    archivePrefix = "arXiv",
    primaryClass = "nucl-th",
    journal = "Phys. Rev. C",
    volume = "93",
    pages = "024002",
    year = "2016"
}

@article{PavonValderrama:2005wv,
    author = "Pavón Valderrama, M. and Ruiz Arriola, E.",
    title = "{Renormalization of NN interaction with chiral two pion exchange potential. central phases and the deuteron}",
    eprint = "nucl-th/0506047",
    archivePrefix = "arXiv",
    journal = "Phys. Rev. C",
    volume = "74",
    pages = "054001",
    year = "2006"
}

@article{PavonValderrama:2005uj,
    author = "Pavón Valderrama, M. and Ruiz Arriola, E.",
    title = "{Renormalization of NN interaction with chiral two pion exchange potential: Non-central phases}",
    eprint = "nucl-th/0507075",
    archivePrefix = "arXiv",
    journal = "Phys. Rev. C",
    volume = "74",
    pages = "064004",
    year = "2006",
    note = "[Erratum: Phys.Rev.C 75, 059905 (2007)]"
}

@article{Entem:2025siq,
    author = "Entem, David R. and Nieves, Juan and Oller, Jose Antonio",
    title = "{Contact potentials in the presence of a regular finite-range interaction using dimensional regularization and the N/D method}",
    eprint = "2506.01461",
    archivePrefix = "arXiv",
    primaryClass = "nucl-th",
    journal = "Phys. Rev. D",
    volume = "112",
    number = "9",
    pages = "096007",
    year = "2025"
}

@article{Entem:2021kvs,
    author = "Entem, D. R. and Oller, J. A.",
    title = "{Non-perturbative methods for NN singular interactions}",
    eprint = "2103.02069",
    archivePrefix = "arXiv",
    primaryClass = "nucl-th",
    journal = "Eur. Phys. J. ST",
    volume = "230",
    number = "6",
    pages = "1675--1689",
    year = "2021"
}

@article{Chew:1960iv,
    author = "Chew, Geoffrey F. and Mandelstam, Stanley",
    title = "{Theory of low-energy pion pion interactions}",
    journal = "Phys. Rev.",
    volume = "119",
    pages = "467--477",
    year = "1960"
}

@article{Entem:2016ipb,
    author = "Entem, D. R. and Oller, J. A.",
    title = "{The N/D method with non-perturbative left-hand-cut discontinuity and the $^1S_0$ $NN$ partial wave}",
    eprint = "1610.01040",
    archivePrefix = "arXiv",
    primaryClass = "nucl-th",
    journal = "Phys. Lett. B",
    volume = "773",
    pages = "498--504",
    year = "2017"
}

@article{Oller:2018zts,
    author = "Oller, J. A. and Entem, D. R.",
    title = "{The exact discontinuity of a partial wave along the left-hand cut and the exact $N/D$ method in non-relativistic scattering}",
    eprint = "1810.12242",
    archivePrefix = "arXiv",
    primaryClass = "hep-ph",
    journal = "Annals Phys.",
    volume = "411",
    pages = "167965",
    year = "2019"
}

@ARTICLE{Birse:2005um,
  author = {Birse, Michael C.},
  title = {Power counting with one-pion exchange},
  journal = {Phys. Rev. C},
  year = {2006},
  volume = {74},
  pages = {014003},
  month = {Jul},
  issue = {1},
  numpages = {14},
  publisher = {American Physical Society},
}

@ARTICLE{Epelbaum_2018,
  author = {Epelbaum, E. and Gasparyan, A. M. and Gegelia, J. and Meissner, Ulf-G.},
  title = {How (not) to renormalize integral equations with singular potentials
	in effective field theory},
  journal = {The European Physical Journal A},
  year = {2018},
  volume = {54},
  number = {11},
}

@ARTICLE{Epelbaum:2020maf,
  author = {Epelbaum, E. and Gasparyan, A. M. and Gegelia, J. and Meißner, Ulf-G.
	and Ren, X.-L.},
  title = {How to renormalize integral equations with singular potentials in
	effective field theory},
  journal = {The European Physical Journal A},
  year = {2020},
  volume = {56},
  pages = {152},
}

@ARTICLE{Epelbaum:2009sd,
  author = {Epelbaum, E. and Gegelia, J.},
  title = {Regularization, renormalization and “peratization” in effective field
	theory for two nucleons},
  journal = {The European Physical Journal A},
  year = {2009},
  volume = {41},
  pages = {341},
}

@ARTICLE{Epelbaum:2006pt,
  author = {Epelbaum, E. and Meißner, Ulf-G.},
  title = {On the Renormalization of the One–Pion Exchange Potential and the
	Consistency of Weinberg’s Power Counting},
  journal = {Few-Body Systems},
  year = {2013},
  volume = {54},
  pages = {2175},
}

@article{Kaplan:1998we,
    author = "Kaplan, David B. and Savage, Martin J. and Wise, Mark B.",
    title = "{Two nucleon systems from effective field theory}",
    eprint = "nucl-th/9802075",
    archivePrefix = "arXiv",
    reportNumber = "DOE-ER-40561-357, INT-98-00-5, NT-UW-98-08, CALT-68-2161",
    journal = "Nucl. Phys. B",
    volume = "534",
    pages = "329--355",
    year = "1998"
}

@article{Kaplan:1998tg,
    author = "Kaplan, David B. and Savage, Martin J. and Wise, Mark B.",
    title = "{A New expansion for nucleon-nucleon interactions}",
    eprint = "nucl-th/9801034",
    archivePrefix = "arXiv",
    reportNumber = "DOE-ER-40561-352, INT-97-00-189, NT-UW-98-05, CALT-68-2155",
    journal = "Phys. Lett. B",
    volume = "424",
    pages = "390--396",
    year = "1998"
}

@article{Kaplan:1996xu,
    author = "Kaplan, David B. and Savage, Martin J. and Wise, Mark B.",
    title = "{Nucleon - nucleon scattering from effective field theory}",
    eprint = "nucl-th/9605002",
    archivePrefix = "arXiv",
    reportNumber = "DOE-ER-40561-257, INT-96-00-125, UW-PT-96-06, CMU-HEP-96-06, DOE-ER-40862-117, CALT-68-2047",
    journal = "Nucl. Phys. B",
    volume = "478",
    pages = "629--659",
    year = "1996"
}

@article{Long:2011xw,
    author = "Long, Bingwei and Yang, C. J.",
    title = "{Renormalizing Chiral Nuclear Forces: Triplet Channels}",
    eprint = "1111.3993",
    archivePrefix = "arXiv",
    primaryClass = "nucl-th",
    reportNumber = "JLAB-THY-11-1464, INT-PUB-11-038",
    journal = "Phys. Rev. C",
    volume = "85",
    pages = "034002",
    year = "2012"
}

@article{Long:2012ve,
    author = "Long, Bingwei and Yang, C. J.",
    title = "{Short-range nuclear forces in singlet channels}",
    eprint = "1202.4053",
    archivePrefix = "arXiv",
    primaryClass = "nucl-th",
    reportNumber = "JLAB-THY-12-1495, INT-PUB-12-001",
    journal = "Phys. Rev. C",
    volume = "86",
    pages = "024001",
    year = "2012"
}

@article{Long:2011qx,
    author = "Long, Bingwei and Yang, C. J.",
    title = "{Renormalizing chiral nuclear forces: a case study of 3P0}",
    eprint = "1108.0985",
    archivePrefix = "arXiv",
    primaryClass = "nucl-th",
    reportNumber = "JLAB-THY-11-1401",
    journal = "Phys. Rev. C",
    volume = "84",
    pages = "057001",
    year = "2011"
}

@ARTICLE{Machleidt:2010kb,
  author = {R Machleidt and D R Entem},
  title = {Nuclear forces from chiral {EFT}: the unfinished business},
  journal = {Journal of Physics G: Nuclear and Particle Physics},
  year = {2010},
  volume = {37},
  pages = {064041},
  number = {6},
  month = {may},
  publisher = {{IOP} Publishing},
}

@ARTICLE{Marji:2013uia,
  author = {Marji, E. and Canul, A. and MacPherson, Q. and Winzer, R. and Zeoli,
	Ch. and Entem, D. R. and Machleidt, R.},
  title = {Nonperturbative renormalization of the chiral nucleon-nucleon interaction
	up to next-to-next-to-leading order},
  journal = {Phys. Rev. C},
  year = {2013},
  volume = {88},
  pages = {054002},
  month = {Nov},
  issue = {5},
  numpages = {10},
  publisher = {American Physical Society},
}

@ARTICLE{PhysRevC.72.054006,
  author = {Nogga, A. and Timmermans, R. G. E. and van Kolck, U.},
  title = {Renormalization of one-pion exchange and power counting},
  journal = {Phys. Rev. C},
  year = {2005},
  volume = {72},
  pages = {054006},
  month = {Nov},
  issue = {5},
  numpages = {16},
  publisher = {American Physical Society},
}

@ARTICLE{PhysRevC.53.2086,
  author = {Ord\'o\~nez, C. and Ray, L. and van Kolck, U.},
  title = {Two-nucleon potential from chiral Lagrangians},
  journal = {Phys. Rev. C},
  year = {1996},
  volume = {53},
  pages = {2086--2105},
  month = {May},
  issue = {5},
  numpages = {0},
  publisher = {American Physical Society},
}

@ARTICLE{PhysRevLett.72.1982,
  author = {Ord\'o\~nez, C. and Ray, L. and van Kolck, U.},
  title = {Nucleon-nucleon potential from an effective chiral Lagrangian},
  journal = {Phys. Rev. Lett.},
  year = {1994},
  volume = {72},
  pages = {1982--1985},
  month = {Mar},
  issue = {13},
  numpages = {0},
  publisher = {American Physical Society},
}

@ARTICLE{ORDONEZ1992459,
  author = {C. Ordóñez and U. van Kolck},
  title = {Chiral lagrangians and nuclear forces},
  journal = {Physics Letters B},
  year = {1992},
  volume = {291},
  pages = {459 - 464},
  number = {4},
  issn = {0370-2693},
}

@article{NavarroPerez:2013usk,
    author = "Navarro P{\'e}rez, R. and Amaro, J. E. and Ruiz Arriola, E.",
    title = "{Partial Wave Analysis of Nucleon-Nucleon Scattering below pion production threshold}",
    eprint = "1304.0895",
    archivePrefix = "arXiv",
    primaryClass = "nucl-th",
    journal = "Phys. Rev. C",
    volume = "88",
    pages = "024002",
    year = "2013",
    note = "[Erratum: Phys.Rev.C 88, 069902 (2013)]"
}

@ARTICLE{PhysRevC.88.064002,
  author = {P\'erez, R. Navarro and Amaro, J. E. and Arriola, E. Ruiz},
  title = {Coarse-grained potential analysis of neutron-proton and proton-proton
	scattering below the pion production threshold},
  journal = {Phys. Rev. C},
  year = {2013},
  volume = {88},
  pages = {064002},
  month = {Dec},
  issue = {6},
  numpages = {31},
  publisher = {American Physical Society},
}

@ARTICLE{Valderrama:2009ei,
  author = {Pav\'on Valderrama, M.},
  title = {Perturbative renormalizability of chiral two-pion exchange in nucleon-nucleon
	scattering},
  journal = {Phys. Rev. C},
  year = {2011},
  volume = {83},
  pages = {024003},
  month = {Feb},
  issue = {2},
  numpages = {11},
  publisher = {American Physical Society},
}

@ARTICLE{PavonValderrama:2011fcz,
  author = {Pav\'on Valderrama, M.},
  title = {Perturbative renormalizability of chiral two-pion exchange in nucleon-nucleon
	scattering: $P$ and $D$ waves},
  journal = {Phys. Rev. C},
  year = {2011},
  volume = {84},
  pages = {064002},
  month = {Dec},
  issue = {6},
  numpages = {23},
  publisher = {American Physical Society},
}

@article{Weinberg:1991um,
    author = "Weinberg, Steven",
    title = "{Effective chiral Lagrangians for nucleon - pion interactions and nuclear forces}",
    reportNumber = "UTTG-03-91",
    journal = "Nucl. Phys. B",
    volume = "363",
    pages = "3--18",
    year = "1991"
}

@article{Weinberg:1990rz,
    author = "Weinberg, Steven",
    title = "{Nuclear forces from chiral Lagrangians}",
    reportNumber = "UTTG-31-90",
    journal = "Phys. Lett. B",
    volume = "251",
    pages = "288--292",
    year = "1990"
}

@ARTICLE{Zeoli:2012bi,
  author = {Zeoli, Ch. and Machleidt, R. and Entem, D. R.},
  title = {{Infinite-cutoff renormalization of the chiral nucleon-nucleon interaction
	at $N^{3}LO$}},
  journal = {Few Body Syst.},
  year = {2013},
  volume = {54},
  pages = {2191--2205},
  archiveprefix = {arXiv},
  eprint = {1208.2657},
  primaryclass = {nucl-th}
}

@article{Case:1950an,
    author = "Case, K. M.",
    title = "{Singular potentials}",
    journal = "Phys. Rev.",
    volume = "80",
    pages = "797--806",
    year = "1950"
}

@article{Frank:1971xx,
    author = "Frank, W. and Land, D. J. and Spector, R. M.",
    title = "{Singular potentials}",
    journal = "Rev. Mod. Phys.",
    volume = "43",
    pages = "36--98",
    year = "1971"
}

@article{PavonValderrama:2005gu,
    author = "Pavón Valderrama, M. and Ruiz Arriola, E.",
    title = "{Renormalization of the deuteron with one pion exchange}",
    eprint = "nucl-th/0504067",
    archivePrefix = "arXiv",
    journal = "Phys. Rev. C",
    volume = "72",
    pages = "054002",
    year = "2005"
}

@article{Beane:2000fi,
    author = "Beane, Silas R. and Savage, Martin J.",
    title = "{Rearranging pionless effective field theory}",
    eprint = "nucl-th/0011067",
    archivePrefix = "arXiv",
    reportNumber = "NT-UW-00-028, JLAB-THY-00-63, NT@UW-00-028",
    journal = "Nucl. Phys. A",
    volume = "694",
    pages = "511--524",
    year = "2001"
}

@article{10.1214/aos/1176344552,
author = {B. Efron},
title = {{Bootstrap Methods: Another Look at the Jackknife}},
volume = {7},
journal = {The Annals of Statistics},
number = {1},
publisher = {Institute of Mathematical Statistics},
pages = {1 -- 26},
year = {1979},
}

\end{document}